\documentclass[structabstract]{aa}  
\usepackage{graphicx}
\usepackage[usenames,dvipsnames]{color}

\usepackage{txfonts}
\usepackage{longtable}
\usepackage{supertabular}
\usepackage{natbib}
\defcitealias{Fadda+08}{Paper 0}
\usepackage{color}
\begin{document}

\newcommand{\mic}{$\mu$m$\,$}
\newcommand{\mica}{$\mu$m}
\newcommand{\lir}{$\rm{L}_{\rm{IR}} \,$}
\newcommand{\lira}{$\rm{L}_{\rm{IR}}$}
\newcommand{\tsfr}{$\Sigma$(SFR)$\,$}
\newcommand{\tsfra}{$\Sigma$(SFR)}
\newcommand{\tsfrm}{\Sigma$(SFR)$/M \,}
\newcommand{\tsfrma}{\Sigma$(SFR)$/M}
\newcommand{\sfrd}{{\rho}_{\rm{SFR}}}

\title{The evolution of galaxy star formation activity in massive haloes
\thanks{Herschel is an ESA space observatory with science instruments provided by 
European-led Principal Investigator consortia and with important 
participation from NASA.}}
\author{P. Popesso\inst{1,2}, A. Biviano\inst{3}, Finoguenov\inst{2}, D. Wilman\inst{2}, M. Salvato\inst{2}, B. Magnelli\inst{2}, C. Gruppioni\inst{4}, F. Pozzi\inst{5 }, G. Rodighiero\inst{6}, F. Ziparo\inst{2}, S. Berta\inst{2}, D. Elbaz\inst{7}, M.~Dickinson\inst{8},D. Lutz\inst{2}
\and B. Altieri\inst{9}
\and H. Aussel\inst{7}
\and A. Cimatti\inst{6}
\and D. Fadda\inst{10}
\and O. Ilbert\inst{11}
\and E. Le Floch\inst{7}
\and R. Nordon\inst{2}
\and A. Poglitsch\inst{2}
\and C.K. Xu\inst{12}
}
\offprints{Paola Popesso, popesso@mpe.mpg.de}

\institute{Excellence Cluster Universe, Boltzmannstr. 2, D-85748 Garching.
\and Max-Planck-Institut f\"{u}r Extraterrestrische Physik (MPE), Postfach 1312, 85741 Garching, Germany.
\and INAF/Osservatorio Astronomico di Trieste, via G.B. Tiepolo 11, Trieste (Italy) I-34143 
\and INAF-Osservatorio Astronomico di Bologna, via Ranzani 1, I-40127 Bologna, Italy.
\and Dipartimento di Astronomia, Universit{\`a} di Bologna, Via Ranzani 1, 40127 Bologna, Italy.
\and Dipartimento di Astronomia, Universit{\`a} di Padova, Vicolo dell'Osservatorio 3, 35122 Padova, Italy.
\and Laboratoire AIM, CEA/DSM-CNRS-Universit{\'e} Paris Diderot, IRFU/Service
 d'Astrophysique,  B\^at.709, CEA-Saclay, 91191 Gif-sur-Yvette Cedex, France.
\and  National Optical Astronomy Observatory, 950 North Cherry Avenue, Tucson, AZ 85719, USA
\and Herschel Science Centre, European Space Astronomy Centre, ESA, Villanueva de la Ca\~nada, 28691 Madrid, Spain
\and NASA Herschel Science Center, Caltech 100-22, Pasadena, CA 91125, USA
\and Institute for Astronomy 2680 Woodlawn Drive Honolulu, HI 96822-1897, USA 
\and IPAC, Caltech 100-22, Pasadena, CA 91125
}

\date{Received / Accepted}

\abstract{There is now a large consensus that the current epoch of the
  cosmic star formation history (CSFH) is dominated by low mass
  galaxies while the most active phase, between redshifts 1 and 2, is
  dominated by more massive galaxies, which
  evolve  more quickly.}{Massive galaxies tend to inhabit very massive haloes, such
  as galaxy groups and clusters. We aim to understand whether the
  observed ``galaxy downsizing'' could be interpreted as a ``halo
  downsizing'', whereas the most massive haloes, and their galaxy
  populations, evolve more rapidly than the haloes with lower mass.}{We
  studied the contribution to the CSFH of galaxies inhabiting
  group-sized haloes. This is done through the study of the evolution
  of the infra-red (IR) luminosity function of group galaxies from
  redshift 0 to redshift $\sim$ 1.6. We used a sample of 39 X-ray-selected groups in the Extended Chandra
  Deep Field South (ECDFS), the Chandra Deep Field North (CDFN), and
  the COSMOS field, where the deepest available mid- and far-IR
  surveys have been conducted with Spitzer MIPS and with the
  Photodetector Array Camera and Spectrometer (PACS) on board the
  Herschel satellite.}{Groups at low redshift lack the brightest,
  rarest, and most star forming IR-emitting galaxies observed in the
  field. Their IR-emitting galaxies contribute $\le 10\%$ of the
  comoving volume density of the whole IR galaxy population in the local Universe. At
  redshift $\gtrsim 1$, the most IR-luminous galaxies (LIRGs and ULIRGs) are
mainly located in groups, and this is consistent with a
  reversal of the star formation rate (SFR) vs.  density anti-correlation
  observed in the nearby Universe. At these redshifts, group galaxies
  contribute 60-80\% of the CSFH, i.e. much more
  than at lower redshifts. Below $z \sim 1$, the comoving number and
  SFR densities of IR-emitting galaxies in groups decline
  significantly faster than those of all IR-emitting galaxies.}{Our
  results are consistent with a ``halo downsizing'' scenario and
  highlight the significant role of ``environment'' quenching in
  shaping the CSFH.}

\keywords{Galaxies: star formation - Galaxies: clusters: general - Galaxies: evolution - Galaxies: starburst}

\titlerunning{The evolution of galaxy star formation activity in massive haloes}
\authorrunning{Popesso et al.}

\maketitle

\section{Introduction}
\label{s:intro}

 One of the most fundamental correlations between the properties of
galaxies in the local Universe is the so-called morphology-density
relation.  Since the early work of \citet{Dressler80}, a plethora of
studies utilizing multi-wavelength tracers of activity have shown that
late type star-forming galaxies favour low-density regimes in the local
Universe \citep[e.g.][]{Gomez+03}. In particular, the cores of massive
galaxy clusters are galaxy graveyards of massive spheroidal systems
dominated by old stellar populations.  Much of the current debate
centres on whether the relation arises early on during the formation
of the object, or whether it is caused by environment-driven
evolution. However, as we approach the epoch when the quiescent
behemoths should be forming the bulk of their stars at $z \gtrsim 1.5$
\citep[e.g.][]{Rettura+10}, the relation between star formation (SF)
activity and environment should progressively reverse. \citet{Elbaz+07} and \citet{Cooper+08} observe the reversal of the
star formation rate (SFR) vs. density relation already at $z\sim1$ in
the GOODS and the DEEP2 fields, respectively. Using {\it{Herschel}}
PACS data, \citet{Popesso+11} show that the reversal is mainly
observed in high-mass galaxies and is due to a higher fraction of
active galactic nuclei (AGN), which exhibit slightly higher SFR than
galaxies of the same stellar mass \citep{Santini+12}. On the other
hand, \citet{Feruglio+10} find no reversal in the COSMOS field and
argue that the reversal, if any, must occur at $z\sim 2$. Similarly,
\citet{Ziparo+14} find that the local anti-correlation tends to
flatten towards high redshift rather than reversing.

On a related topic, there is now a large consensus that the cosmic
star formation history (CSFH) peaks at increasingly higher redshifts
for galaxies of higher stellar mass at redshift zero. The
star formation activity of low-mass galaxies (stellar mass $\leq
10^{10}$ $M_{\odot}$) peaks at redshift $z \sim 0.2$, whereas that of
more massive galaxies (stellar mass$ \geq 10^{11}$ $M_{\odot}$)
monotonically declines from $z \sim 0.5-1$ or higher \citep[up to $z \sim
2$ for stellar mass$ >10^{12}$ $M_{\odot}$]{heavens+04,gruppioni+13}. This monotonic decline
leads to a decrease of an order of magnitude in the SFR density of the
Universe after z $\sim$ 1.  Highly star-forming galaxies such as the
luminous infrared galaxies (LIRGs) are rare in the local Universe, but
are the main contributors to the CSFH at $z \sim 1-3$ \citep{LeFloch+05,Perez-Gonzalez+2005,Caputi+07,reddy+08,magnelli+09,magnelli+11,magnelli+13}. The most powerful starburst galaxies,
such as the ultraluminous infrared galaxies (ULIRGs) and the sub-mm
galaxies undergo the fastest evolution, dominating the CSFH only at $z
\sim 2-3$ and disappearing, then, by redshift $\sim$ 0 \citep{cowie+04}.  Most massive galaxies seem to have formed their stars
early in cosmic history, and their contribution to the CSFH was
significantly greater at higher redshifts through a very powerful
phase of SF activity (LIRGs, ULIRGs and sub-mm galaxies). Low-mass
galaxies seem to have formed much later, and they dominate the present
epoch through a mild and steady SF activity. This phenomenology is
generally referred to as "galaxy downsizing". The evidence that more
massive galaxies tend to reside in more massive haloes sets a clear
link between the galaxy SF activity evolution and their
environment. The "galaxy downsizing" scenario could therefore be
interpreted in terms of a ``halo downsizing'' scenario as highlighted
in \cite{NvdBD06} and \citet{popesso+12}.

The most straighforward way to probe whether there is a reversal of
the SFR-density relation in the distant Universe and what the
contribution of galaxies in massive haloes is to the CSFH is to study the
evolution of the SFR density of galaxies in such haloes. In most
galaxies, the bulk of UV photons, emitted by young, massive stars, is
absorbed by dust and re-emitted at infrared (IR) wavelengths
\citep[see][]{Kennicutt98}.  For this reason the IR luminosity is a
very robust indicator of the bolometric output from young stars and
therefore a good proxy for the galaxy star formation rate \citep[SFR,
  e.g.][]{buat+02,bell+03}. As a consequence, the evolution of the
galaxy IR luminosity function (LF) provides a direct measure of the
evolution of the galaxy SFR distribution. 

In the local Universe the bulk of the total stellar mass is contained
in galaxy groups with total mass greater than $10^{12.5}$ $M_{\odot}$.
The fraction of stellar mass in the more massive galaxy clusters is
negligible because these are rare objects \citep{eke+05}, and
group-sized haloes are the most common high-mass haloes for a galaxy to
inhabit. For this reason, we study the evolution of the IR LF of
galaxies in groups, and compare it to that of more isolated field
galaxies.  While the IR LF of field galaxies and its evolution are
relatively well known up to $z \sim 2.5-3$ 
\citep{Caputi+07,magnelli+09,Gruppioni+11,gruppioni+13,reddy+08}, the
IR LF of galaxies in groups and clusters is still poorly known. 

 Most determinations of galaxy IR LFs in cluster and supercluster
environments have so far been based on Spitzer
data. \citet{Bai+06,Bai+09} analyzed the IR LFs of the rich nearby
clusters Coma and A3266. According to their analysis, the bright end
of the IR LF has a universal form for local rich clusters, and cluster
and field IR LFs have similar values for their characteristic
luminosities. \citet{Bai+09} compare the average IR LFs of these two
nearby clusters and two distant ($z \sim 0.8$) systems and conclude
that there is a redshift evolution of both the characteristic
luminosity and the normalization of the LF such that higher-z clusters
contain more and brighter IR galaxies. Other studies find considerable
variance in the IR LF in galaxy clusters, and this might be related to
the presence of substructures \citep{Chung+10,Biviano+11}.

Much less is known about the evolution of the IR LF in dark matter
haloes of lower mass, such as the galaxy groups. \citet{Tran+09}
determine the IR LFs in a rich galaxy cluster and four galaxy groups
at $z \sim 0.35$. There are four times more galaxies with a high SFR in the groups than in the cluster, or
equivalently, the group IR LF has an excess at the bright end relative
to the cluster IR LF.  On the basis of this result, \citet{Chung+10}
interpret the excess of bright IR sources in the IR LF of the Bullet
cluster ($z \sim 0.3$) as being due to the galaxy population in an
infalling group (the ``bullet'' itself). \citet{Biviano+11} find that
the IR LF of galaxies in a $z \sim 0.2$ large-scale filament has a
bright-end excess compared to the IR LF of its neighbouring
cluster. Given that the physical conditions (density, velocity
dispersions) are similar in filaments and poor groups, their result
appears to be consistent with \citet{Tran+09}'s.

In this paper we analyse the evolution of the group galaxy IR LF from
redshift 0 to redshift $z \sim 1.6$ by using the newest and deepest
available mid- and far-IR surveys conducted with {\it{Spitzer}} MIPS
and with the most recent Photodetector Array Camera and Spectrometer
(PACS) on board the {\it{Herschel}} satellite, on the major blank
fields such as the Extended Chandra Deep Fields South (ECDFS), the
Chandra Deep Field North (CDFN) and the COSMOS field. All these fields
are part of the largest GT and KT Herschel Programmes conducted with
PACS: the PACS Evolutionary Probe \citep{Lutz+11} and the
GOODS-Herschel Program \citep{Elbaz+11}. In addition, the blank fields
considered in this work are observed extensively in the X-ray with
$Chandra$ and $XMM-Newton$. The ECDFS, CDFN, and COSMOS fields are also
the site of extensive spectroscopic campaigns that have led to excellent
spectroscopic coverage. This is essential for correctly identifying group members.  We use the evolution of the group IR LF to study
the SFR distribution of group galaxies and to measure their
contribution to the CSFH.

The paper is structured as follows. In Sect. \ref{s:data} we describe
our dataset. In Sect. \ref{s:lf} we determine the IR LF in groups. In
Sect.  \ref{field_comp} we compare the IR LF of group galaxies with
the IR LF of the total population. In Sect. \ref{csfh} we analyse the
contribution of group galaxies to the CSFH. In
Sect. \ref{s:discussion} we compare our results with existing models
of galaxy formation and evolution. In Sect. \ref{s:conclusion} we
summarize our results and draw our conclusions.  We adopt H$_0=70$
km~s$^{-1}$~Mpc$^{-1}$, $\Omega_m=0.3$, $\Omega_{\Lambda}=0.7$
throughout this paper.

\section{The data set}
\label{s:data}

\subsection{Infrared and spectroscopic data}
\label{s:specphot}

We use the deepest available {\it{Spitzer}} MIPS 24 $\mu$m and PACS
100 and 160 $\mu$m datasets for all the fields we consider in our
analysis.  For COSMOS these come from from the public Spitzer 24 \mic
\citep{LeFloch+09,Sanders+07} and PEP PACS 100 and 160 \mic data
\citep{Lutz+11,magnelli+13}.  Both {\it{Spitzer}} MIPS 24 and PEP
source catalogues are obtained by applying prior extraction as
described in \cite{magnelli+09}. In short, IRAC and MIPS 24~$\mu$m
source positions are used to detect and extract MIPS and PACS sources,
respectively. This is feasible since extremely deep IRAC and MIPS 24~$\mu$m
observations are available for the COSMOS field
\citep{scoville+07}. The source extraction is based on a PSF-fitting
technique, presented in detail in \cite{magnelli+09}.

The association between 24 \mic and PACS sources, at 100 and 160 $\mu$m, with their optical
counterparts, taken from the optical catalogue of \citet{Capak+07} is done via a maximum likelihood method
\citep[see][for details]{Lutz+11}. The photometric sources were
cross-matched in coordinates with the sources for which a
high-confidence spectroscopic redshift is available. For this purpose
we use the public catalogues of spectroscopic redshifts complemented
with other unpublished data. This catalogue includes redshifts from
either SDSS or the public zCOSMOS-bright data acquired using VLT/VIMOS
\citep{Lilly+07,Lilly+09} complemented with Keck/DEIMOS (PIs:
Scoville, Capak, Salvato, Sanders, Kartaltepe), Magellan/IMACS
\citep{Trump+07}, and MMT \citep{Prescott+06} spectroscopic redshifts.

\begin{table}
\caption{Properties of the PEP fields}
\centering
\begin{tabular}{l c c c c c}
\hline
Field    & Band         & Eff. Area             & 3$\sigma$ \\
                &       &                       & mJy   \\
\hline
GOODS-N &       24 $\mu$m       & 187 arcmin$^2$        & 0.02  \\
GOODS-N &       100 $\mu$m      & 187 arcmin$^2$        & 3.0   \\
GOODS-N &       160 $\mu$m      & 187 arcmin$^2$        & 5.7   \\
\hline
GOODS-S &       24 $\mu$m       & 187 arcmin$^2$        & 0.02  \\
GOODS-S &       70 $\mu$m       & 187 arcmin$^2$        & 1.1   \\
GOODS-S &       100 $\mu$m      & 187 arcmin$^2$        & 0.7   \\
GOODS-S &       160 $\mu$m      & 187 arcmin$^2$        & 1.2   \\
\hline
ECDFS &         24 $\mu$m       & 0.25 deg$^2$          & 0.05  \\
ECDFS &         100 $\mu$m      & 0.25 deg$^2$          & 3.9   \\
ECDFS &         160 $\mu$m      & 0.25 deg$^2$          & 7.5   \\
\hline
COSMOS &        24 $\mu$m       & 2.04 deg$^2$          & 0.06  \\
COSMOS &        100 $\mu$m      & 2.04 deg$^2$          & 5.0   \\
COSMOS &        160 $\mu$m      & 2.04 deg$^2$          & 10.2  \\
\hline
\end{tabular}
\tablefoot{The first column gives the name of the field, the second
  column the MIPS and PACS band in which the field is observed, the third
  column the effective area covered, and the fourth column the
  3$\sigma$ detection limit in mJy.}
\label{tab:irdata}
\end{table}

In the ECDFS and GOODS regions, the deepest available MIR and FIR
data are provided by the {\it{Spitzer}} MIPS 24 $\mu$m Fidel Program
\citep{magnelli+09} and by the combination of the PACS PEP
\citep{Lutz+11} and GOODS-Herschel \citep{Elbaz+11} surveys at 70, 100,
and 160 $\mu$m. The GOODS Herschel survey covers a smaller central
portion of the entire GOODS-S and GOODS-N regions. Recently the PEP
and the GOODS-H teams have combined the two sets of PACS observations to
obtain the deepest ever available PACS maps \citep{magnelli+13} of
both fields. The more extended ECDFS area has been observed in the PEP
survey as well, down to a higher flux limit. As for the COSMOS
catalogues, the 24 \mic and PACS sources in the ECDFS and GOODS fields
are associated to their optical counterparts (provided by the
\citealt{cardamone+10} catalogue for ECDFS, the \citealt{Santini+09}
catalogue for GOODS-S, and the dedicated PEP multi-wavelength
\citealt{Berta+10} catalogue for GOODS-N) via a maximum likelihood
method \citep[see][for details]{Lutz+11}. The photometric sources were
cross-matched in coordinates with the sources with a high-confidence
spectroscopic redshift. The redshift
compilation in the ECDFS and GOODS-S region is obtained by
complementing the spectroscopic redshifts contained in the
\citet{cardamone+10} catalogue with all new publicly available
spectroscopic redshifts, such as the one of \cite{silverman+10} and
the Arizona ECDFS Environment Survey \citep[ACES,][]{cooper+12}. We
clean the new compilation from redshift duplications for the same
source by matching the Cardamone et al. (2010) catalogue with the
\citet{cooper+12} and the \citet{silverman+10} catalogues within
$1\arcsec$ and by keeping the most accurate ${\rm z_{spec}}$ entry
(smaller error and/or higher quality flag) in case of multiple
entries. With the same procedure we include the very high-quality
redshifts of the GMASS survey \citep{Cimatti+08}. The spectroscopic
redshift compilation for the GOODS-N region is taken from
\citet{barger+08}.

Limiting fluxes in the mid- and far-IR for all fields used in this
work are given in Table \ref{tab:irdata}.

\begin{figure}
\begin{center}
\includegraphics[width=0.49\textwidth,angle=0]{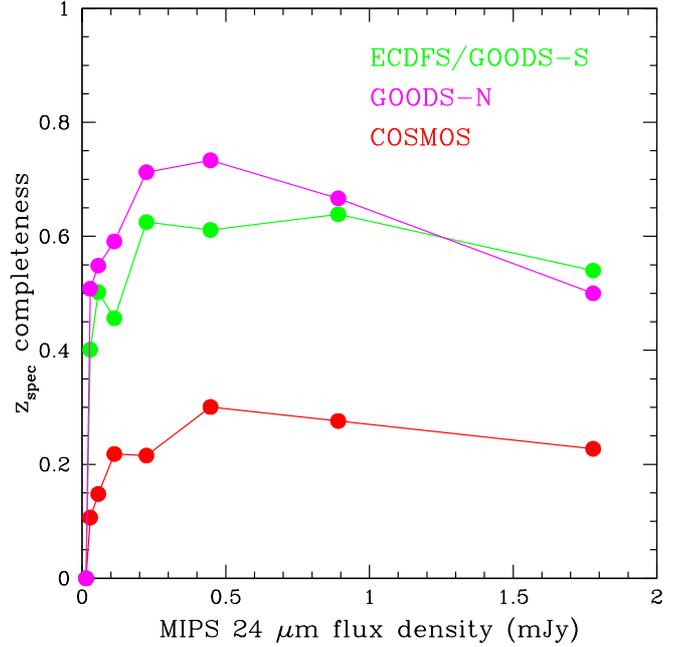} 
\caption{Mean spectroscopic completeness in the
  {\it{Spitzer}} MIPS 24 $\mu$m band across all the areas of the
  ECDFS, GOODS-N, and COSMOS field.}
\label{histogram}
\end{center}
\end{figure}

\subsection{The X-ray selected galaxy group sample}
\label{s:sample}

All the blank fields considered in our analysis have been observed
extensively in the X-ray with $Chandra$ and XMM-$Newton$. To create a
statistically significant sample of galaxy groups, we combined the
X-ray selected group sample of Popesso et al. (2012) and a newly
created X-ray selected group sample of the ECDFS \citep{ziparo+13}. The
sample described in \citet{popesso+12} comprises the X-ray selected
COSMOS group sample of Finoguenov et al. (in preparation) and the
X-ray-detected groups of the GOODS fields. We replace, in particular,
the sample of groups detected in GOODS-S with the sample of groups
from the new catalogue of \citet{ziparo+13} extracted in the larger area
of the ECDFS. The data reduction of the X-ray XMM and Chandra maps of
COSMOS, ECDFS, and GOODS-N were performed in a consistent way and
the initial X-ray group catalogues created according to the
same extended emission extraction procedure (\citealt{finoguenov+09},
Finoguenov et al. in prep.). In short, the point sources were removed
from the X-ray maps. The resulting ``residual'' image is then used
to identify extended emission with at least 4$\sigma$ significance
with respect to the background.

As in \citet{popesso+12} for COSMOS and GOODS-N, \citet{ziparo+13} selected a clean subsample
of groups in the ECDFS catalog with clear spectroscopic redshift
identification along the line of sight, with at least ten members, the
minimum required for a meaningful dynamical analysis, and without
close companions, allowing for a clear definition of the spectroscopic
members. This selection leads to 22 groups in the ECDFS out of the
initial 50 X-ray-selected groups. 

We stress that the imposition of a minimum of ten spectroscopic members is required for a secure velocity dispersion measurement, hence a secure membership definition. This selection does not lead to a bias towards rich systems in our case. Indeed, there is no magnitude or stellar mass limit imposed on the required ten members. Thus, the very high spectroscopic completeneness, in particular of GOODS-N and ECDFS \citep[see][for details]{Popesso+09,cooper+12}, leads to selecting faint and very low-mass galaxy groups. Thus, if the group richness is defined as the number of galaxies brighter than a fixed absolute magnitude limit or more massive than a stellar mass limit, our sample covers a very broad range of richness values, consistent with the scatter observed in the X-ray luminosity-richness relation studied in \citet{Rykoff+12}. In a forthcoming paper (Erfanianfar et al. in prep.), we will extend the current sample to groups with fewer members.

\begin{figure}
\begin{center}
\includegraphics[width=0.49\textwidth,angle=0]{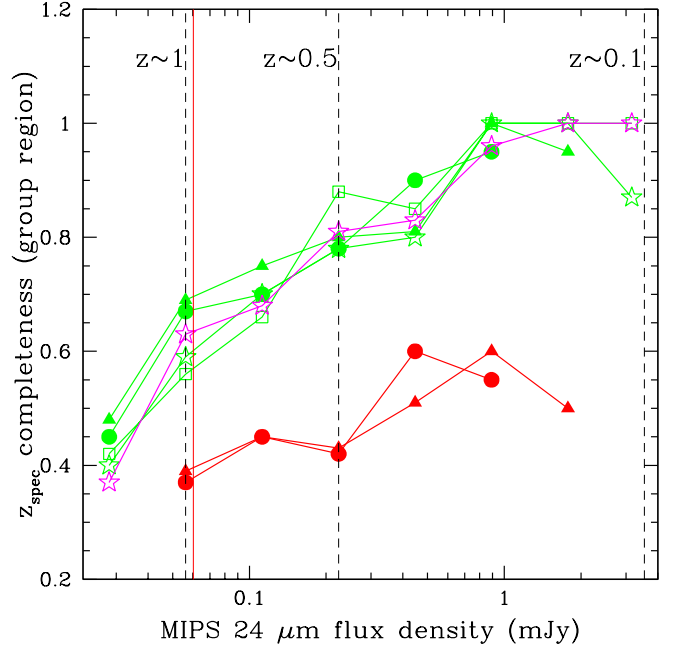} 
\caption{ Mean spectroscopic completeness in the
  {\it{Spitzer}} MIPS 24 $\mu$m band in the group regions as a function of the redshifts. The mean completeness is estimated for groups in different redshift bins and in different fields. The green sysmbols show the mean completeness for groups in ECDFS, magenta symbols show the mean completeness for groups in  GOODS-N, and red symbols show the one for the COSMOS groups. They are grouped in three redshift bins: $z < 0.4$ (filled circles), $0.4 < z < 0.8$ (filled triangles), and $0.8 < z < 1.2$ (stars). The empy squares show the completeness in the region of the \citet{Kurk+08} structure at $z \sim 1.6$ in ECDFS. The mean completeness of the groups in each redshift bin is estimated in the group region within an annulus of $3\times r_{200}$ from the group centre. To guide the eye, the black dashed lines in the figure show the 24 $\mu$m flux corresponding to the LIRG limit ($L_{IR}=10^{11}L_{\odot}$) at low ($z\sim 0.1$), intermediate ($z\sim 0.5$), and high ($z\sim 1$) redshift. The solid red line shows the 60 $\mu$Jy limit used to estimate the mean group completeness. }
\label{histogram1}
\end{center}
\end{figure}

In all fields the total mass of the groups is
derived from their X-ray luminosity ($L_X$) by using the
$L_X-M_{200}$\footnote{The mass $M_{200}$ is the mass enclosed within
  a sphere of radius $r_{200}$, where $r_{200}$ is the radius where
  the mean mass overdensity of the group is 200 times the critical
  density of the Universe at the group mean redshift. } relation of
\citet{Leauthaud+10}. We also impose a mass cut at $M_{200} <
2{\times}10^{14}$ $M_{\odot}$ to avoid including massive clusters,
whose galaxy population could follow a different evolutionary path
from that of groups, as shown in \citet{popesso+12}.

As a result of these selections, our group sample comprises 27 COSMOS
groups at $0 < z < 0.8$, 22 ECDFS groups at $0 < z < 1.0$, two groups
identified in the GOODS-N region at $z \sim 0.85$ and $z \sim 1.05$,
and the GOODS-S group identified by \citet{Kurk+08} at $z \sim
1.6$. This structure was initially optically detected through the
presence of an over-density of [OII] line emitters by
\citet{vanzella+06} and, then, as an over-density of elliptical
galaxies by \citet{Kurk+08} in the GMASS survey. \citet{Tanaka+13}
detected it as an X-ray group candidate in the ECDFS \citep[see
  also][]{ziparo+13}.  The analysis of this system offers the unique
opportunity to attempt to constrain the group IR LF at a very high
redshift. Unlike many other systems at similar redshifts
\citep[e.g.][]{papovich+10}, this structure does not suffer from the
heavy spectroscopic bias against star forming galaxies thanks to the
spectroscopic selection of red and massive galaxies, and its spectroscopic
completeness is indeed high even among IR-emitting galaxies.

We restrict our sample further by selecting only those groups that reach
a spectroscopic completeness in our deepest IR band, the
{\it{Spitzer}} 24 $\mu$m band, of 60\% down to 60 $\mu$Jy.  This flux
detection threshold is reached in all fields at the 3$\sigma$ level or
higher. 


As shown if Fig. \ref{histogram1}, we also check for possible biases due to the spectroscopic selection function of the different fields. In particular, we check for any redshift dependence. For this purpose we estimate the spectroscopic completeness in an
alternative way by using the most accurate photometric redshifts available in the considered fields. We use the photometric redshifts of \citealt{cardamone+10} in ECDFS, the one of \citealt{Berta+10} for GOODS-N and the $z_{phot}$ of \citealt{Ilbert+10} for the COSMOS field. We estimate the number of group member photometric candidates ($N_{z_{phot}}$) as the number of MIPS detected sources in the group region (within an annulus of $3\times r_{200}$ from the group centre) and with $z_{phot}$ within $5\times \sigma_{z_{phot}}$ from the group mean redshift, where $\sigma_{z_{phot}}$ is the photometric redshift uncertainty as reported in the mentioned papers. The completeness is, then, estimated as the subsample of such candidates with spectroscopic redshift and the total number $N_{z_{phot}}$. Figure \ref{histogram1} shows the mean completeness for groups in different redshift bins: $z < 0.4$, $0.4 < z <0.8$, and $0.8< z< 1.2$.  We also show the completeness estimated with this procedure for the  \citet{Kurk+08} structure at $z\sim 1.6$. The result does not change even if we consider a smaller ($3\times \sigma_{z_{phot}}$) or larger ($10\times \sigma_{z_{phot}}$) photometric redshift interval for selecting the group member candidates. It is evident that there is no redshift dependence of the spectroscopic completeness. The main difference arises from the different mean spectroscopic completeness available in the different fields as already shown in Fig \ref{histogram}. In the ECDFS and GOODS-N fields, the spectroscopic selection captures the totality of the bright MIPS group candidates, and it ensures a very high coverage down to the 60 $\mu$Jy limit. This is because all the spectroscopic selection functions are usually biased in favour of emission line galaxies and because the majority of the IR galaxies are part of this class. Thus, our requirement of a 60\% spectroscopic completeness down to 60 $\mu$Jy does not remove any of the groups in the ECDFS
and GOODS-N.  In the COSMOS field, instead,  the mean completeness is lower at any MIPS flux, independently of the group redshift. For this reason, out of
the initial 27 COSMOS groups, only 14 systems lie in a region of the
sky with sufficient spectroscopic coverage to fulfil the selection
criteria. 

After this further selection, our final group sample
comprises 39 groups.  These are used to build composite IR LFs in
different redshift bins: $0 < z < 0.4$ (15 groups), $0.4 < z < 0.8$
(17 groups), $0.8 < z < 1.2$ (6 groups), and the $z\sim 1.6$ group.

\begin{figure}
\begin{center}
\includegraphics[width=0.49\textwidth,angle=0]{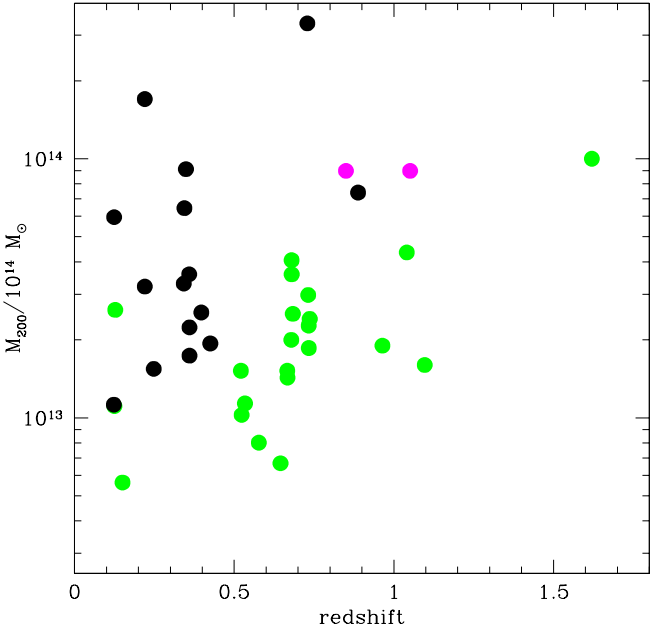} 
\caption{Group masses $M_{200}$ vs. mean redshifts.}
\label{mass_redshift}
\end{center}
\end{figure}

In Fig. \ref{mass_redshift} we show the group masses ($M_{200}$)
vs. their mean redshifts.  The mass-redshift distribution is rather uniform, limiting the bias againt low-mass systems at high redshift.

\subsection{The identification of group members}
\label{s:members}

The group membership is based on the {\it Clean} algorithm of
\citet{Mamon+13}. After selecting the main cluster
peak in redshift space by the method of weighted gaps, the algorithm
estimates the cluster velocity dispersion using the galaxies in the
selected peak. This is then used to evaluate the virial velocity based
on assumed models for the mass and velocity anisotropy profiles. These
models with the estimated virial velocity are then used to predict the
line-of-sight velocity dispersion of the system as a function of
system-centric radius, $\sigma_{\rm{los}}(R)$.  Any galaxy having a
rest-frame velocity within $\pm 2.7 \sigma_{\rm{los}}(R)$ at its
system-centric radial distance $R$, is selected as a group member. 
We use the X-ray surface-brightness peaks as centres of the X-ray
detected systems.  The group members are used to re-compute the total
cluster velocity dispersion, hence its virial velocity, and the
procedure is iterated until convergence.  The value of the virial
velocity obtained at the last iteration of the {\it Clean} algorithm
is used to evaluate the system dynamical mass.

As already mentioned in Popesso et al. (2012), the dynamical and X-ray
mass estimates are in good agreement in the COSMOS field. As discussed
in \citet{ziparo+13}, we note much less agreement for the newly
defined (E)CDFS group sample, where the dynamical masses are on
average higher than the X-ray masses. This could come from the ECDFS groups being on average much more distant than COSMOS groups,
and this is only partially explained by the deeper X-ray exposure in
the ECDFS field with respect to the COSMOS field. In the following we
nevertheless use the masses derived from X-ray luminosities for all
systems, since unlike dynamical masses, they do not suffer
from projection effects, which may be considerable when the number of
spectroscopic members is low, as in our sample.  Only for the $z\sim
1.6$ group do we use the mass obtained by the {\it Clean} algorithm,
because the X-ray luminosity is poorly estimated owing to the very low flux close to the detection level \citep{Tanaka+13}.

\subsection{Bolometric IR luminosity}
\label{s:sfr}
We use the main sequence (MS) and starburst
(SB) templates of \citet{Elbaz+11} to search for the best fit to
the spectral energy distribution (SED) of our group galaxies,
defined by the
PACS (70, 100, and 160\,$\mu$m) fluxes, when available, and by the
24 $\mu$m fluxes.
When the 24 $\mu$m flux is the only one available (i.e. for
undetected PACS sources), we adopt the MS template, since this
provides the best fit to the spectral energy distributions (SEDs) of
most (80\%) sources with both PACS and MIPS 24 $\mu$m detections
\citep[see for a more detailed discussion][]{ziparo+13}. We compute
the IR luminosities ($L_{IR}$) by integrating the best-fit templates in the
range 8-1000\,$\mu$m. 
In principle, using the MS template for the sources with
only 24~$\mu$m fluxes could
cause an underestimation of the extrapolated $L_{IR}$,
in particular for high-redshift or for off-sequence
sources, because of the higher PAHs emission of the MS template
\citep{Elbaz+11,nordon+10}. However, \citet{ziparo+13} have shown that
for the sources with PACS and 24 $\mu$m data,
the $L_{IR}$ estimated with the best-fit
templates are in good agreement with those
estimated using only the
24 $\mu$m flux and the MS template
($L_{IR}^{24} $), with only a slight
underestimation (10\%) at $z\gtrsim 1.7$ and/or $L_{IR}^{24} >
10^{11.7}\ L_\odot$.

\section{The galaxy group IR LF}

\label{s:lf}

\subsection{The composite LF: method}

 Galaxy groups host a relatively small number of (star forming)
galaxies each. As a result, the low statistics prevent us from studying the
individual group LFs. The most straightforward way to overcome this
problem is to consider the average LF of a statistical sample of
groups or, equivalently, the composite IR LF in groups. The most widely
used method of estimating a composite LF is the one of
\citet{colless+89}. In this method, the group galaxies are summed in
IR luminosity bins, and the sums are scaled by the richness of the
parent groups,
\begin{equation} 
N_{gj}=\frac{N_{g0}}{m_j}\sum_i{\frac{N_{ij}}{N_{i0}}}, 
\label{e:colless}
\end{equation} 
where $N_{gj}$ is the number of galaxies in the $j\rm{th}$ IR
luminosity bin of the composite LF, $N_{ij}$ is the number in the
$j\rm{th}$ bin of the $i\rm{th}$ IR LF in groups, $N_{i0}$  the
normalization used for the $i\rm{th}$ IR LF in groups (number of group
member brighter than a fixed luminosity), $m_j$  the number of
groups contributing to the $j\rm{th}$ bin, and $N_{g0}$ is the sum of
all the normalizations:
\begin{equation} 
N_{g0}=\sum_i{{N_{i0}}}.  
\label{e:ng0}
\end{equation}

It is easy to note that in the Colless (1989) prescriptions, the $j\rm
{th}$ bin of the composite LF represents just the mean fraction of
galaxies, with respect to the normalization region, of all the groups
contributing to the $j\rm {th}$ bin. In other words,
Eq. \ref{e:colless} provides the mean fractional distribution of
galaxy luminosity, multiplied by an arbitrary normalizaton, $N_{g0}$,
which is just the sum of all the normalizations of the systems
involved in the estimate. To obtain a composite LF with physically
meaningful normalization, we rescale the mean fractional luminosity
distribution,
\begin{equation} 
f_{gj}=\frac{1}{m_j}\sum_i{\frac{N_{ij}}{N_{i0}}} \, ,
\label{e:colless}
\end{equation} 
by the mean group richness, which is the mean number of galaxies
brighter than a given $L_{IR}$ value,
\begin{equation} 
N_{g0_new}=\frac{\sum_i{{N_{i0}}}}{N_{groups}} \, ,
\label{e:ng0}
\end{equation}
where $N_{groups}$ is the number of groups considered for the estimate
of the composite IR LF. The limiting $L_{IR}$ value is set by the
limit reached at the upper boundary of any given redshift bin, namely
$\log L_{IR}/L_{\odot}=9, 10, 11,$ and 11.3, in the redshift bins
0--0.4, 0.4--0.8, 0.8--1.2, and at $z=1.6$, respectively.  In any
redshift bin, groups contribute to the composite LF only down to the
limiting $L_{IR}$ they are sampled to; that is, only the lowest $z$ groups
in any redshift bin contribute down to the the bin-limiting
luminosity.

Since we do not have redshifts for all galaxies in the group fields,
we must correct for spectroscopic incompleteness. In principle, one
should apply this correction to each IR luminosity bin by following the
same method as \citet{depropris+03} for the $b$ band cluster LF. They
estimate the spectroscopic incompleteness correction per apparent
magnitude bin (corresponding to the appropriate absolute magnitude
bin) for each system as the ratio between the number of all galaxies
and of the galaxies with spectroscopic redshifts, in that magnitude
bin. In our case this estimate is complicated by the fact that
$L_{IR}$ is not derived from a single mid or far-IR band but from SED
fitting. There is therefore not a one-to-one relation between luminosities
and observed fluxes. We then adopt the following approach. We assume
that the redshift determinations are unbiased with respect to group
membership, and this assumption is justified by the high spatial
homogeneity of the spectroscopic coverage of our fields \citep[see for
  instance][for the ECDFS]{cooper+12}. We then take as reference
photometric band the mid-IR {\it{Spitzer}} MIPS 24 $\mu$m band,
since it is the deepest IR band in our fields. Following \citet{depropris+03},
we estimate the
correction for incompleteness in the region of each group (within $2
\times r_{200}$ from the group X-ray centre) per bin of MIPS 24 $\mu$m flux
by assigning the following weights to each group member,
\begin{equation}
w_{s,k}=N_{k}/N_{k,spec},
\label{weight}
\end{equation}
where $N_k$ is the number of galaxies in the $k$-th 24 $\mu$m flux
bin, and $N_{k,spec}$  the number of galaxies in that same bin with
spectroscopic determination. If all galaxies in the considered flux
bin have measured redshifts, then $w_{s,k}=1$, otherwise $w_{s,k} >
1$, and the galaxies with $z_{spec}$ also account for those without
measured redshift. The mean value of $w_{s,k} $ in our sample is
$1.5\pm 0.2$, and the maximum is 2.3.  To also take the
photometric incompleteness in the faintest MIPS 24 $\mu$m flux bins into account,
we multiply the weight $w_{s,k}$ by the weight $w_{p,k}$ that is
defined as the inverse of the completeness per flux bin, estimated as
described in \citet{Lutz+11}. These photometric weights are
$w_{p,k} \approx 1.70$, 1.25, 1.00 for sources with fluxes at the 3, 5, 8 ${\sigma}$ detection
levels, respectively, corresponding to 60\%, 80\%, and 100\%
completeness, respectively.  The final weight assigned to each
galaxy is given by $w_k=w_{s,k}{\times} w_{p,k}$. The number of
galaxies in the $i\rm{th}$ group and within the $j\rm{th}$ IR
luminosity bin is then given by
\begin{equation}
N_{ij}=\sum_{\rm{members}}{w_k} \, ,
\end{equation}
where only spectroscopic members of the given group and in the given
luminosity bin are considered. In the same way the normalization
$N_{i0} $ of the individual IR LF in groups is obtained as the sum of
$w_k$ of the group members with $L_{IR}$ brighter than the IR
luminosity limit.

Following \citet{depropris+03}, the formal error on $N_{gj}$ is obtained by
propagating the errors on $N_{ij}$ and $N_{i0}$, which are both given
by the sum in quadrature of the error of the weight $w_k$ of the
contributing group members. In Eq. \ref{weight} ,$N_k$ is a Poisson
variable, since it is drawn from an ideal (infinite) distribution, and
$N_{k,spec}$ is a binomial random variable, the number of `successes'
(redshift determinations) in $n$ `trials' (number of spectroscopic
targets) with probability of success, $p$, given by the success rate
of the spectroscopic campaign. We estimate this success rate to be equal to
0.7-0.8 in the GOODS and ECDFS and COSMOS regions given the estimates
reported by the major spectroscopic campaigns conducted in these
fields in the redshift range considered here
\citep{barger+08,Popesso+09,balestra+10,cooper+12,Lilly+09}. Therefore
the error on $w_k$ is given by
\begin{equation}
{{\delta^2 w_k} \over {w^2_k}}={{\sigma^2 (N_{k})} \over N^2_{k}}+
{{\sigma^2 (N_{k,spec})} \over N^2_{k,spec}} ~,
\end{equation}
where we neglect the contribution to the error of the photometric
incompleteness weights $w_{p,k}$, which turns out to be extremely
stable in the simulations performed in all fields \citep{Lutz+11,
  magnelli+13}. If we consider that the Poissonian error $\sigma^2
(N_{k})=N_{k}$ and the standard deviation of the binomial random
variable $N_{k,spec}$ is
$\sigma^2(N_{k,spec})=n{\times}p{\times}(1-p)$ according to the
standard binomial error expression, where $n=N_{k,spec}/p$, the
previous equation simplifies to
\begin{equation}
{{\delta^2 w_{k, member}} \over {w^2_{k, member}}}
           = {1\over N_k}+{(1-p)\over N_{k,spec}} ~.
\label{e:werr}
\end{equation}

\subsection{The composite LF: results}

\begin{figure}
\begin{center}
\includegraphics[width=0.49\textwidth]{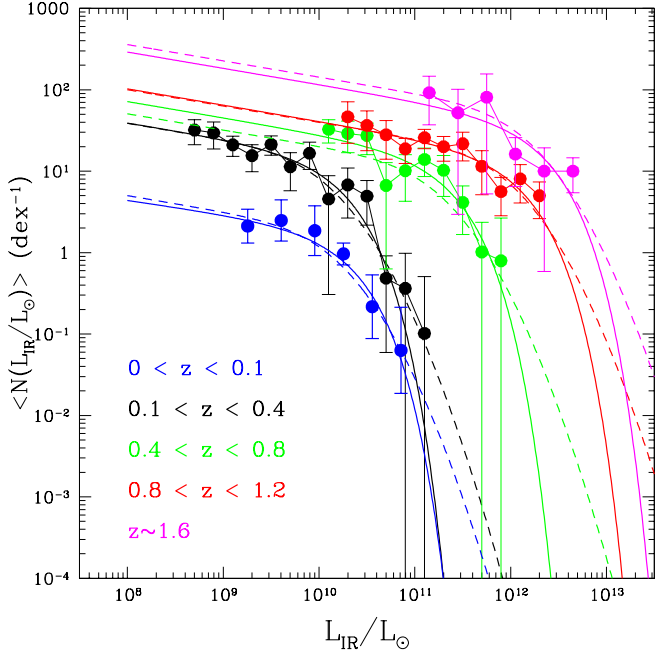} 
\caption{IR LFs in galaxy groups estimated within $2{\times}r_{200}$
  in different redshift bins: $0 < z < 0.4$ (black points), $0.4 < z <
  0.8$ (green points), $0.8 < z < 1.2$ (red points), and at $z\sim
  1.6$ (magenta points; this is the IR LF of the structure of
  \citet{Kurk+08} ). Error bars are 1 $\sigma$. The IR LF in groups in the
  lowest redshift bin at $z<0.1$ (blue points) is obtained by
  averaging the the IR LF in groups obtained in three different group halo
  mass bins by \citet{Guo+14}.  The solid (resp. dashed) lines,
  colour-coded as the points of the IR LF, indicate the best fit
  Schechter \citep[resp. modified Schechter, see][]{saunders+90}
  functions.}
\label{irlf}
\end{center}
\end{figure}

The composite IR LF for group galaxies is shown in Fig. \ref{irlf} in
the four redshift bins. In the highest redshift bin at $z\sim1.6$, the
IR LF is that of the \citet{Kurk+08} structure.  We
estimate the composite IR LF of groups within $r_{200}$ and
$2{\times}r_{200}$. Figure \ref{irlf} shows, in particular, the IR LF
obtained within the larger physical aperture. The LF estimated within
the two apertures are consistent. We only notice that the lower
statistics observed within the smaller radius ($r_{200}$) leads to a
slightly noisier LF, while the higher statistics obtained within the
largest physical aperture ($2{\times}r_{200}$) allow the best-fit parameters to be better
constrained.

Our IR LFs do not sample the $z<0.1$ redshift range. For this we use
the lowest redshift determination of the IR LF of the \citet{Robotham+11} groups by \citet{Guo+14}. This LF is based on the 250
$\mu$m luminosity ($L_{\nu}(250 {\mu}m)$ ) of the H-ATLAS SPIRE survey of a 135 $\rm{deg}^2$ region.  In addition, we
use also the LFs derived by \citet{Guo+14} in the other redshift
bins (0.1--0.2, 0.2--0.3, and 0.3--0.4) to check the consistency with
our LF determination in the redshift bin 0.1--0.4. For a meaningful
comparison, we average the LFs obtained by \citet{Guo+14} in
several group mass bins from $10^{12.5}$ to $10^{13}$ $M_{\odot}$. We
use as error bars the dispersion of the LF in any $L_{\nu}(250
{\mu}m)$ luminosity bin. As a last
step we use equation 2 of \citet{Guo+14} to transform the
group $L_{\nu}(250 {\mu}m)$ LF into the IR LF in groups.

We use three different fitting functions to find the best fit: the
Schechter function, the modified Schechter function of \citet{saunders+90}, and a double power law similar to the one used by
\citet{sanders+03} for local star-forming galaxies,
\begin{equation}
{\phi}(L)dL=\left(\frac{\phi^*}{L^*}\right)\left({\frac{L}{L^*}}\right)^{\alpha}e^{-\frac{L}{L^*}}dL \, ,
\end{equation}
\begin{equation}
{\phi}(L)dLog(L)=\phi^*\left({\frac{L}{L^*}}\right)^{1+\alpha}e^{-\frac{1}{2\sigma^2}log_{10}^2\left(\frac{L}{L^*}\right)}dLog(L) \, ,
\end{equation}
\begin{eqnarray}
{\phi}(L)dL={\Phi_a}L^{\alpha}dL\quad if \; {L<L_{knee}}, \nonumber \\
{\phi}(L)dL={\Phi_b}L^{\beta}dL\quad if\; {L>L_{knee}}, \nonumber \\
\mathrm{where:} \; \; {\Phi_a}={\Phi_b}L_{knee}^{\beta-\alpha} \, .
\end{eqnarray}

The double power law provides the worst fits in all cases and will not
be considered in the further analysis. Indeed we do not observe a
clear ``knee'' in the IR LF in groups (see Fig. \ref{irlf}) as observed
instead in the total IR LF
\citep[e.g.][]{sanders+03,magnelli+09,magnelli+11,magnelli+13}. The
group galaxy IR LF shows a smoother decline at high luminosity which
is more consistent with a Schechter or a modified Schechter function
than with a double power law. The double power law predicts too many
bright galaxies at the bright end, especially in the two lowest
redshift bins, where we observe a rather fast decline of the LF at
very high luminosity. The Schecter and modified Schechter fits (both
shown in Fig. \ref{irlf}) are of similar quality. Free parameters of
the fits are the LF normalization and its ``characteristic'' or
``knee'' luminosity $L^*$. The LF slope parameter $\alpha$ is a free
parameter only in the lowest redshift bin ($0.1 < z <0.4$), where we
sample the IR LF to relatively low values of $L_{IR}$. We use this
best-fit value of $\alpha$ for the higher redshift IR LFs, where the
faint end is not well sampled by our data. A similar approach has been
taken by \citet{magnelli+09} and \citet{gruppioni+13}. At $ z < 0.1$
for the IR LF derived from \citet{Guo+14}, the shallower depth of
the H-ATLAS survey does not allow constraining the LF faint end. Thus, in this case we fix $\alpha$ to the value estimated at $0.1 < z <
0.4   $. In the case of the modified Schechter function of \citet{saunders+90}, we fix the $\sigma$ parameter to the value of 0.5 as in
Gruppioni et al. (2013). The best-fit parameters are listed in
Table~\ref{tab:irfits}.

Consistently with the behaviour already observed in the total IR LF at
similar redshifts
\citep[e.g.][]{magnelli+09,magnelli+11,magnelli+13,gruppioni+13}, the
knee luminosity and the normalization of the group galaxy LF increases
with redshift; that is, both the number of star forming galaxies in groups and
their mean IR luminosity increase with redshift.  This reflects the
generally increasing SF activity of the Universe with redshift.

\begin{figure}
\begin{center}
\includegraphics[width=0.49\textwidth]{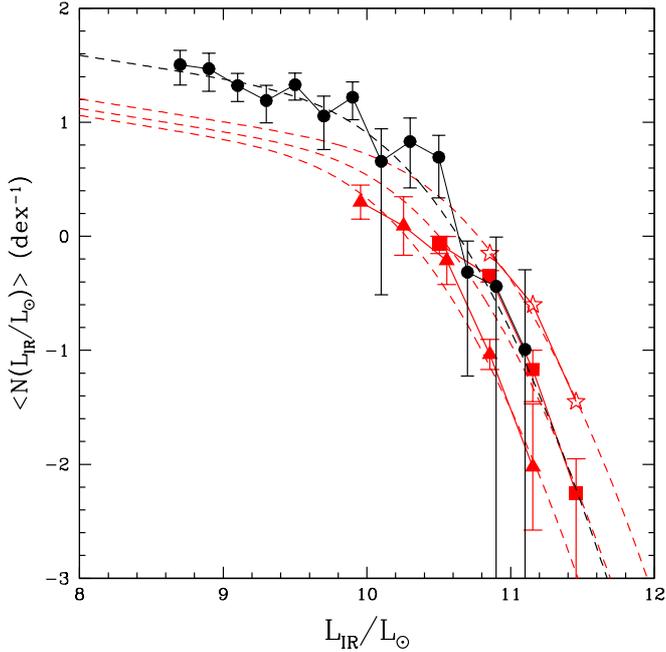} 
\caption{Comparison of the IR LFs in groups (black filled points and solid
  lines) at $0.1 < z < 0.4$ derived in this work and the group LFs of
  \citet{Guo+14} (red symbols and dashed lines) converted into
  IR LF derived at $0.1 < 0 < 0.2$ (triangles), $0.2 < 0 < 0.3$
  (squares), and $0.3 < z < 0.4$ (stars).  For clarity we show here
  only the best fit derived with the modified Schechter function of
  \citet{saunders+90} .}
\label{guo}
\end{center}
\end{figure}

The comparison of our LFs with those derived by \citet{Guo+14} is
shown in Fig. \ref{guo}. Our determination of the group LFs reaches
fainter luminosities, so we can only compare the bright end of the
group LFs. While there is agreement between the Guo et al. and our LFs at
the very bright end, though our error bar are very large, the normalizations are different, and the Guo et
al. (2014) groups contain, on average, a lower number of IR-emitting galaxies than the groups observed in this work. It is hard to
tell whether this is an effect of the blending problems of the very large SPIRE PSF of the H-ATLAS maps or of the different group-selection technique of \citet{Robotham+11}. In the former case, the large SPIRE PSF at 250 ${\mu}$ of $\sim$ 18 arcsec could lead to blending problems in crowded regions, such as groups and clusters. Two or a few relatively faint sources that are closer than the PSF FWHM are therefore identified as a single brighter source. This would subtract sources from the faint end towards the bright end. In the latter case, instead, the optical selection could lead to selecting low-richness groups for a given halo mass that are usually undetected in the X-rays. This would lead to a lower mean group richness, hence to a lower LF normalization.

\begin{figure}
\begin{center}
\includegraphics[width=0.49\textwidth]{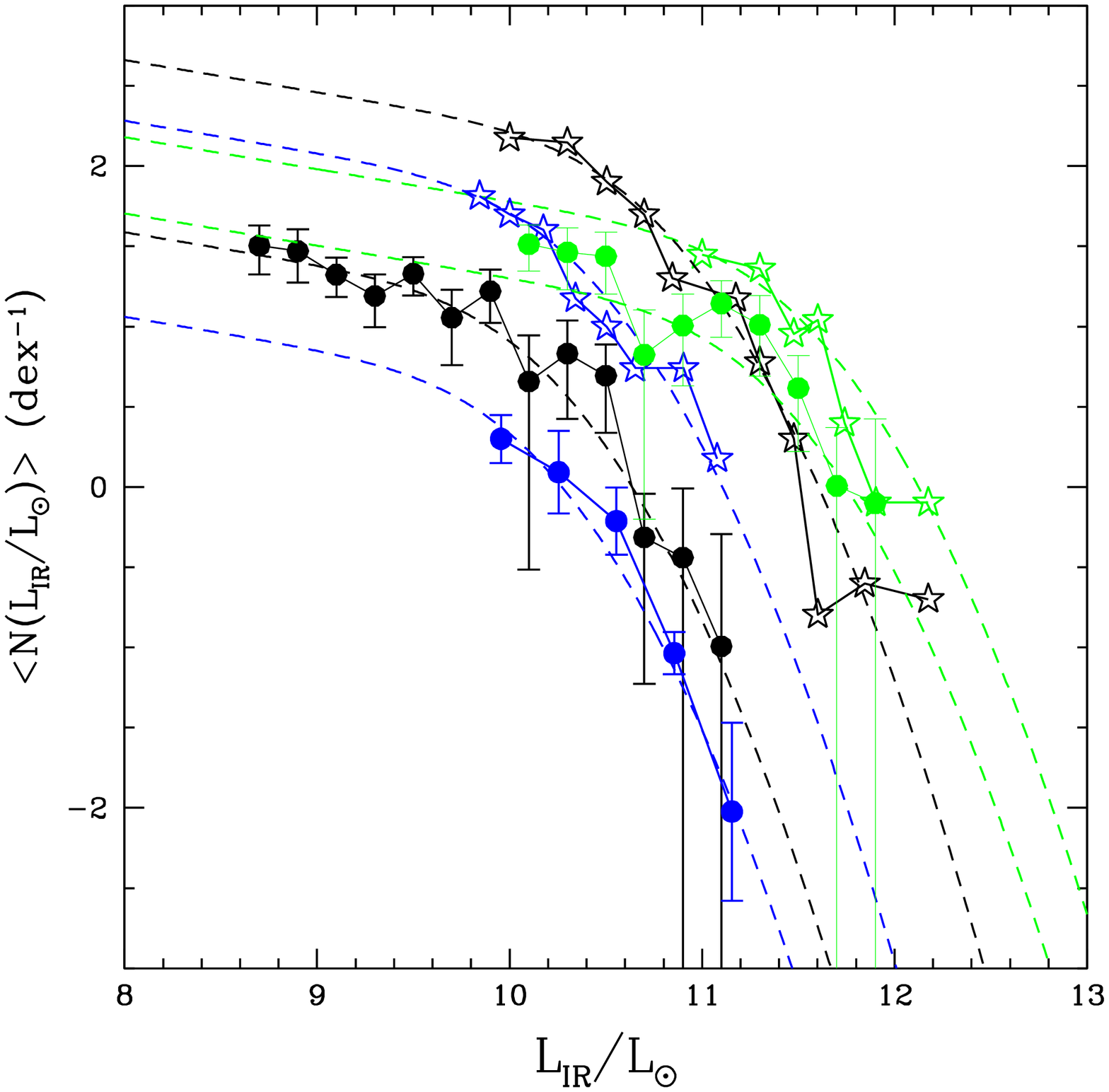} 
\caption{Comparison of the IR LFs in groups (filled points) with the
  IR LF in clusters (stars). The colour code is the same as in Fig. \ref{irlf}: blue
  for the nearby groups of \citet{Guo+14} and the Shapley superclusters
  \citep{Haines+10} at $z < 0.1$, black for $0.1 < z < 0.4$ groups
  and $0.15 < z <0.3$ LoCuSS clusters \citep{Haines+13}, and green
  for $0.4 < z <0.8$ groups and for the $0.6 <z < 0.8$ rich cluster LF
  of \citep{Finn+10}. For clarity we only show here the best fit
  derived with the modified Schechter function of \citet{saunders+90} .}
\label{cl}
\end{center}
\end{figure}

For completeness we also compare the IR LFs in clusters
available in the literature. Figure \ref{cl} shows the comparison of the
\citet{Guo+14}  IR LF in groups at $z < 0.1$ with the IR LF of the
Shapley supercluster studied in Haines et al. (2010), which as shown
in Haines et al. (2013), is consistent with the IR LF of the Coma
cluster and A3266 studied by Bai et al. (2006) and Bai et al. (2009),
respectively. Our IR LF in groups at $0.1 < z < 0.4$ is compared with the
stacked IR LF of 30 clusters observed at $0.15 < z < 0.30$ in the
LoCuSS survey (Haines et al. 2013). Our IR LF in groups at $0.4 < z < 0.8$
is compared with the stacked IR LF of six rich clusters at $0.6 < z <
0.8$ studied in Finn et al. (2010). We fit the cluster LFs with the
Schechter and modified Schechter functions (see
Table~\ref{tab:irfits}).  The main difference between the group and the
cluster LF is obviously the normalization, since the clusters are much
richer, so they contain many more star-forming and IR-emitting
galaxies. However, modulo the different normalization, the knee
luminosity of cluster and group LFs shows a similar evolution
(see also Fig. \ref{comp2}, upper panel).

\begin{figure*}
\begin{center}
\includegraphics[width=0.79\textwidth]{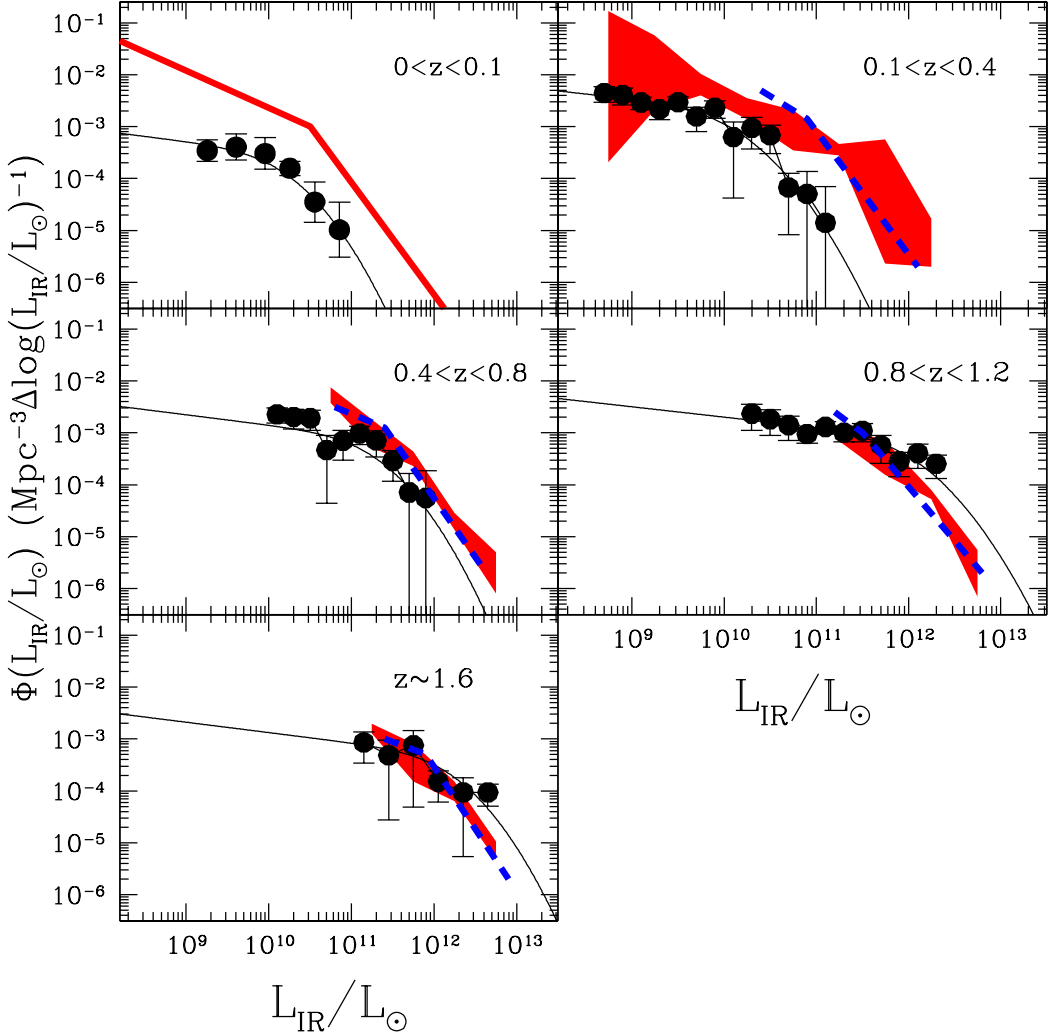} 
\caption{Comparison of IR LF in groups (filled points) and the total IR LF
  (shaded red regions).  The observed IR luminosity function in galaxy
  groups is indicated by the points with error bars. In the first
  redshift bin at $z < 0.1,$ we show the IR LF in groups derived from Guo
  et al. (2014) and the best fit of the field IR LF of Sanders et
  al. (2003). At higher redshift we use the total IR LF of Gruppioni
  et al. (2013, red shaded region) and Magnelli et al. (2013, blue
  dashed line). The IR LF in groups modified Schechter function best fit
  is shown by the solid line in all cases. The different luminosity
  limit of the group LF with respect to the total LF of
  \citet{gruppioni+13} and Magnelli et al. (2013) is because the group LF is also based on MIPS 24 $\mu$m data, while the
  general LF is based only on PACS data.}
\label{comp}
\end{center}
\end{figure*}

\begin{table}
\centering
\caption{Best-fit parameters of the IR LFs}
\begin{tabular}{c c c c}
\hline
\hline
  redshift    & $log(\Phi*)$   &  $\alpha$      & $log(L^*)$    \\
\hline
\hline
\multicolumn{4}{c}{IR LFs of the groups in our sample} \\
\hline
\multicolumn{4}{c}{Schechter function} \\
   $0.1 < z  < 0.4$ & 1.4$\pm$0.05 & -1.19$\pm$0.16 & 10.24$\pm$0.02 \\
  $0.4 <z  < 0.8$  & 2.54$\pm$0.15 & -1.2 & 11.36$\pm$0.05 \\
  $0.8 < z < 1.2$ & 3.29 $\pm$ 0.17 & -1.2 & 12.1$\pm$0.14 \\
  $z \sim 1.6$ & 3.93$\pm$0.5 & -1.2 & 12.34$\pm$0.31 \\
\hline
\multicolumn{4}{c}{modified Schechter function} \\
\hline
 redshift    & $log(\Phi*)$   &  $\alpha$       & $log(L^*)$    \\
   $0.1 < z  < 0.4$ & 1.27$\pm$0.04 & -1.2$\pm$0.21 & 9.57$\pm$0.01 \\
   $0.4 <z  < 0.8$  & 1.15$\pm$0.16 & -1.2 & 10.73$\pm$0.04 \\
   $0.8 < z < 1.2$ & 1.3$\pm$0.16 & -1.2 & 11.46$\pm$0.13 \\
   $z \sim 1.6$ & 1.82$\pm$0.73 &-1.2 & 11.65$\pm$0.22 \\
\hline
\hline
\multicolumn{4}{c}{IR LFs of the groups of \citet{Guo+14}} \\
\hline
\multicolumn{4}{c}{Schechter function} \\
\hline
   $0 < z < 0.1$  & 0.57$\pm$0.05 & -1.17$\pm$0.02 & 10.32$\pm$0.005 \\
   $0.1 < z < 0.2$ & 0.74$\pm$0.08&-1.2 & 10.43$\pm$0.08 \\
   $0.2 < z < 0.3$& 0.84$\pm$0.09 & -1.2 &10.72$\pm$0.11 \\
   $0.3 < z < 0.4$ &1.13$\pm$0.11 & -1.2 &10.89$\pm$0.15 \\
\hline
\multicolumn{4}{c}{modified Schechter  function} \\
\hline
  $0 < z < 0.1$  & 0.37$\pm$0.03 & -1.2$\pm$0.02 & 9.64$\pm$0.003 \\
   $0.1 < z < 0.2$ & 0.75$\pm$0.08&-1.2 & 9.50$\pm$0.04 \\
   $0.2 < z < 0.3$& 0.77$\pm$0.09 & -1.2 &9.72$\pm$0.12 \\
   $0.3 < z < 0.4$ &0.81$\pm$0.15 & -1.2 &9.97$\pm$0.17 \\
\hline
\hline
\multicolumn{4}{c}{IR LFs of clusters} \\
\hline
\multicolumn{4}{c}{Schechter function} \\
\hline
 $z\sim 0.05$ &2.11$\pm$0.04 & -1.2 &10.53$\pm$0.15\\
$0.15 < z < 0.3 $ & 2.79$\pm$0.05 & -1.2 &10.84$\pm$0.17\\ 
$0.6 < z < 0.8$ & 2.97$\pm$0.08 & -1.2 & 11.34$\pm$0.11\\
\hline
\multicolumn{4}{c}{modified Schechter function} \\
\hline
 $z\sim 0.05$ &1.93$\pm$0.04 & -1.2 & 9.74$\pm$0.15 \\
$0.15 < z < 0.3$ & 2.31$\pm$0.04 & -1.2 & 9.99$\pm$0.18\\
$0.6 < z <0.8$ & 1.6$\pm$0.08 & -1.2 & 10.90$\pm$0.09 \\
\hline
\hline
\end{tabular}
\tablefoot{ Best-fit parameters of the Schechter function and of the modified Schechter function of Saunders et al. (1990) for the composite IR LF of our group sample, the local mean IR LF derived from the work of \citet{Guo+14}, and for the cluster IR LF of Haines et al. (2013), respectively.}
\label{tab:irfits}
\end{table}

\section{Comparison with the total IR LF}
\label{field_comp}
In this section we compare the IR LF of our groups with the total IR
LF at similar redshifts. For this we must first evaluate the IR LF of
the group galaxies per unit of comoving volume. To do so, we multiply
our determination of the group IR LF by the comoving number density of
the dark matter haloes in the same mass range as a function of redshift
($\rho_{N_{halo}}(z)$).  (We consider here the
$10^{13-14}$ $M_{\odot}$ halo mass range as the average normalization of
the group IR LF is dominated by groups in this mass range.) We estimate $\rho_{N_{halo}}(z)$ by using the
WMAP9 concordance-model \citep{Hinshaw+13}
prediction of the comoving $dN/dz$ of haloes in the mass range of our group sample. This model reproduces the
observed $\log(N)-\log(S)$ distribution of the deepest X-ray group and
cluster surveys \citep[see e.g.][]{finoguenov+10}. For comparison we
estimate the comoving $dN/dz$ in the same mass bin, also according to
the Planck cosmology based on the SZ Planck number counts
\citep{PC+13}. In this cosmology, the number of groups is
0.14 dex higher, on average, up to $z\sim 1.5$. We caution that the
estimate of the IR LF of the group galaxy population at $z\sim 1.6$ is
only tentative since it is based on one group only, which is relatively
massive and might not be representative of the general group galaxy
population at that redshift.

In Fig. \ref{comp} we show the IR LF of the group galaxy population,
expressed as number of galaxies per unit of comoving volume, together
with the total IR LFs as derived by \citet{gruppioni+13} and
\citet{magnelli+13}. The best-fit faint-end slope we determined for
the $0.1 < z < 0.4$ group LF is in remarkable agreement with the one
obtained for the field LF by Gruppioni er al. (2013), while the LF of
Magnelli et al. (2013) shows a steeper faint end, which is, however,
consistent with the results of \citet{magnelli+09,magnelli+11} based
on deeper Spitzer data.  The shape of the IR LF in groups at $ 0.4 < z <
0.8$ is consistent, in terms of $L^*$ and faint-end slope, with that
of the total IR LF in the same redshift bin, even if the volume
density of IR-emitting group galaxies is $\sim$ 60\% lower than the
volume density of the total population. Indeed, if we re-normalize
the group and the total IR LF to their integral over the same
luminosity region, the two LFs overlap perfectly. This similarity is
not observed in other redshift bins.

At $z< 0.4$, the group IR LF is characterized by a much steeper
cut-off at the bright end than the total IR LF. Groups at this
redshift lack the brightest, rarest, most star-forming IR
galaxies that are instead observed in the field. In addition, the
volume density of the group IR-emitting galaxies is much lower (less
than 10\%) than the volume density of the whole IR galaxy population.

\begin{figure}
\begin{center}
\includegraphics[width=0.49\textwidth]{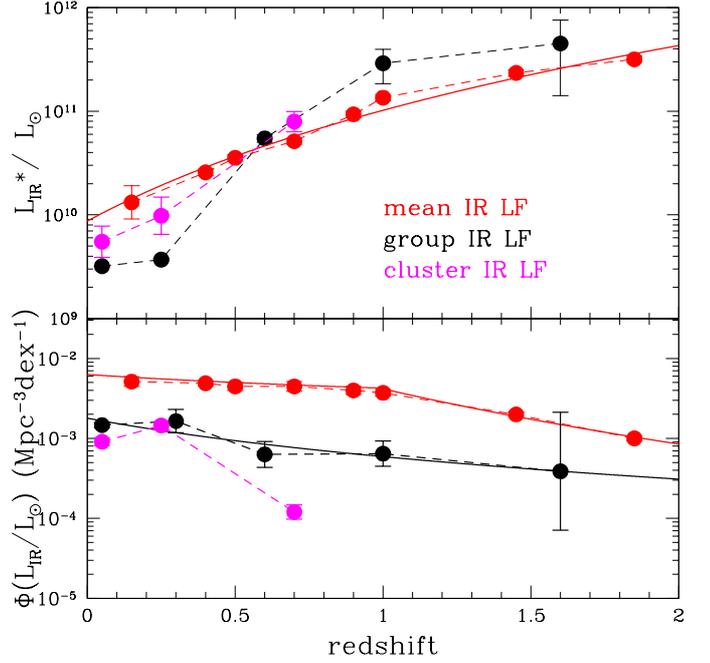}
\caption{Evolution of the best-fit $L^*$ (upper panel) and $\Phi^*$
    (bottom panel) parameters of the modified Schechter of
    \citet{saunders+90} for the IR LF in groups (black points), the total IR LF of \citet[][red points]{gruppioni+13}, and the cluster IR LF
    (magenta points). Error bars are 1 $\sigma$ (when not shown they
    are smaller than the symbol size).}
\label{comp2}
\end{center}
\end{figure}

At $z \sim 1$ the IR LF in groups and in the Kurk et al. (2008)
structure exhibit a slighty brighter knee luminosity than the
total LF (see also the upper panel of Fig. \ref{comp2}). In addition,
the normalization of the LF indicates that the density of IR-emitting
group galaxies is very close to the density of the total galaxy
population. This indicates that at high redshift, the group galaxy
population makes a substantial contribution to the IR emitting
galaxy population as a whole. The slightly higher $L^*$ of the group
galaxy LF indicates a potentially higher mean SFR in groups than in
the field, at least in the star-forming galaxy population. This
would be consistent with a flattening of the SFR-density relation at
this redshift, as found by Ziparo et al. (2014), but also with a
potential reversal of the same relation, though the difference between
the total and group galaxy luminosity distribution is not very
significant ($\sim 2\sigma$, see Fig. \ref{comp2}).

For a more quantitative comparison, we show the
best-fit values of the $L^*$ (upper panel) and ${\Phi}*$ (lower panel)
parameters of the modified Schechter function in Fig. \ref{comp2}, as a function of
redshift, for both the group and the total IR LFs, and for the
IR LF in clusters at the redshifts where these parameters have been
estimated. To obtain the ${\Phi}*$ of the cluster IR LF, we multiplied
the cluster IR LF shown in Fig. \ref{cl} by the number density of
cluster size haloes (at masses over $10^{14}$ $M_{\odot}$) as a
function of redshift. We only consider the IR LF of
\citet{gruppioni+13} here. A direct comparison with the LF of
\citet{magnelli+13} is not possible, since they use a double power law
function that does not provide a good fit to the group LF.

The evolution of the knee luminosity seems to be faster in groups than
for the total galaxy population. An power law does not provide
a good fit to the $L^*$-redshift relation as in Gruppioni et
al. (2013) and Magnelli et al. (2013). Most of the evolution seems to
take place at $z > 0.4$. Below this redshift, both $L^*$ and ${\Phi}^*$
are not evolving significantly. We observe a mild ${\Phi}^*$ evolution
as a function of redshift which is well fitted by a power law
${\Phi}^* \propto z^{-1.6}$.

\begin{figure}
\begin{center}
\includegraphics[width=0.49\textwidth]{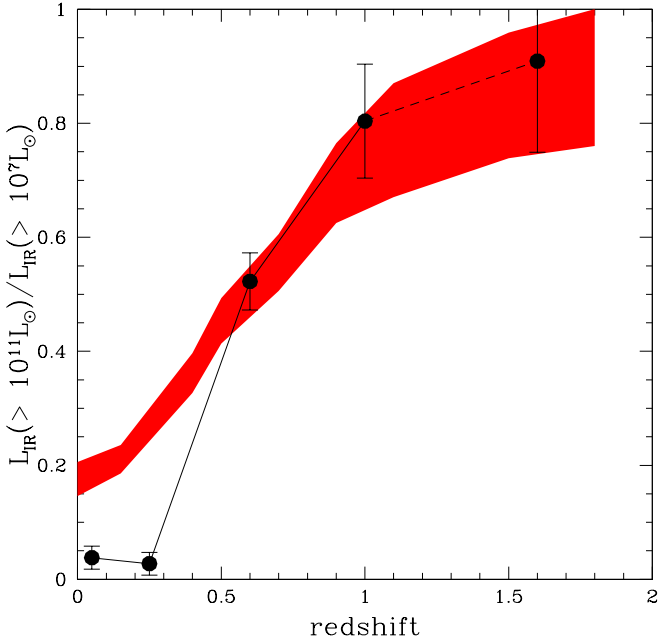}
\caption{Evolution of the fraction of IR
  luminosity due to the LIRGs in groups (black points) and in the
  total IR galaxy population (red shaded region). This is obtained as
  the ratio between the integral of the observed IR LF down to $L_{IR}
  > 10^{11} L_{\odot}$ and the integral estimated down to $L_{IR} >
  10^{7} L_{\odot}$ of the group and total IR LF,
  respectively. The red shaded region is obtained by considering in
  any redshift bin the whole range of possible values, including
  errors, derived from the total IR LF estimated by
  \citet{sanders+03,LeFloch+05,rodighiero+10a,magnelli+09,magnelli+11,magnelli+13,gruppioni+13}. }
\label{comp1}
\end{center}
\end{figure}

To provide a more model-independent comparison, we show in
Fig. \ref{comp1} the fraction of IR luminosity due to LIRGs in groups
and in the total IR galaxy population. This fraction is obtained as
the ratio between the integral of the observed IR LF down to $L_{IR} >
10^{11} L_{\odot}$ and the integral estimated down to $L_{IR} > 10^{7}
L_{\odot}$ of the group and total IR LF, respectively. The two
integrals are estimated by using both the observed LF in the luminosity
range sampled by the observations and the best-fit LF
(modified Schechter function) at fainter luminosities. Up to
$z=1.2$ the integral down to $L_{IR} =10^{11} L_{\odot}$ is
based entirely on the observed LF. The correction of the integral down to
$L_{IR} =10^{7} L_{\odot}$ due to the extrapolation from the best fit
LF is $<10$\%.  At $z > 1$ also the integral down to $L_{IR} =10^{11}
L_{\odot}$ must be corrected with an extrapolation of the best fit
LFs. The correction is however very small (7\% at the limit $L_{IR}
=10^{11} L_{\odot}$, and 15\% down to $L_{IR} =10^{7} L_{\odot}$)
because LIRGs account for most of the IR luminosity.

The contribution of the LIRG population in the total galaxy
population is shown by the red region of Fig. \ref{comp2}. This region
is obtained by considering in any redshift bin the whole range of
possible values, including errors, derived from the total IR LF
estimated by
\citet{sanders+03,LeFloch+05,rodighiero+10a,magnelli+09,magnelli+11,magnelli+13,gruppioni+13}. 
We observe a significantly faster decline
of the LIRG luminosity contribution in groups with respect to the
total galaxy population.  At high redshift ($ z > 0.4$), the
fractional luminosity contribution of LIRGs is similar in groups and
in the total galaxy population. At lower redshift this contribution
is close to zero in galaxy groups, while it is 5-10\% in the total
population.

There have been numerous studies on the origin and evolution of LIRGs,
suggesting that these galaxies -- at least in the local Universe --
are triggered by strong interactions and mergers of gas-rich galaxies
\citep[see the review by][]{sandres_mirabel+96}. The fraction of
mergers among LIRGs increases with IR luminosity and approaches 100\%
for samples of nearby ULIRGs
\citep{sanders+88,kim+95,clements+96,farrah+01,veilleux+02}. At higher
redshift, most of the LIRGs are MS galaxies, and they are generally
not associated to merger activity
\citep{bell+05,Elbaz+07,Elbaz+11}. Thus, our findings seem to suggest
that galaxy groups are not favorite sites for the onset of a strong
merger activity, at least not among gas-rich galaxies and not at
$z<1$.

\section{The contribution of group galaxies to the SFR density of the Universe}
\label{csfh}
\begin{figure}
\begin{center}
\includegraphics[width=0.49\textwidth]{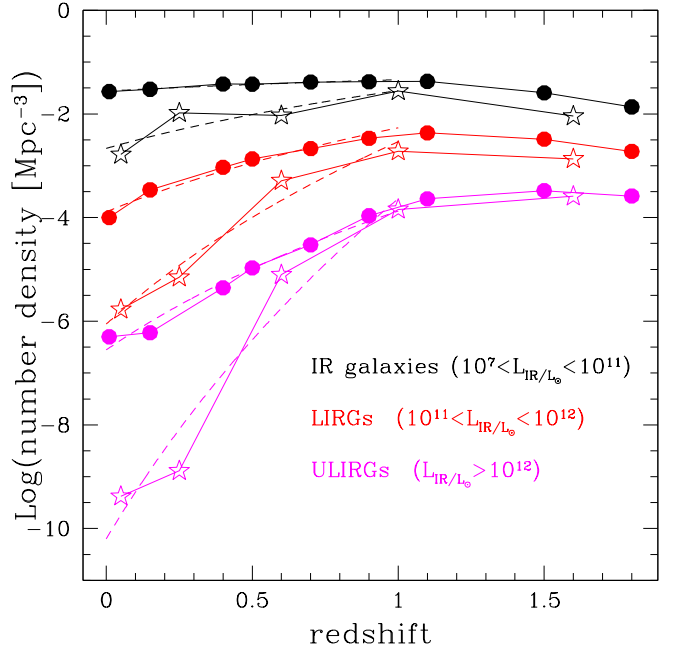}
\caption{Evolution of the comoving number density of faint IR emitting
  galaxies ($10^7 < L_{IR}/L_{\odot} < 10^{11}$, black symbols), LIRGs
  ($10^{11} \leq L_{IR}/L_{\odot} < 10^{12}$, red symbols), and ULIRGs
  ($L_{IR}/L_{\odot} \geq 10^{12}$, magenta symbols) of the whole
  galaxy population (filled points) and of the group galaxy population
  (stars). The comoving number density of the total population is
  taken from Magnelli et al. (2013) at $z > 0.1$ and from Sanders et
  al. (2003) at $z \sim 0$. The dashed lines show the best fit
  relation between comoving number density and redshift,
  $(1+z)^\alpha$ (best-fit values of $\alpha$ are given in the text).}
\label{num_density}
\end{center}
\end{figure}

By integrating the group galaxy IR LFs, we derive the evolution of the
comoving number density (Fig.~\ref{num_density}) of ``faint'' galaxies
(i.e., $10^7\,$L$_{\odot}$$\,<\,$$L_{\rm
  IR}$$\,<\,$$10^{11}\,$L$_{\odot}$), LIRGs (i.e.,
$10^{11}\,$L$_{\odot}$$\,<\,$$L_{\rm
  IR}$$\,<\,$$10^{12}\,$L$_{\odot}$), and ULIRGs (i.e., $L_{\rm
  IR}$$\,>\,$$10^{12}\,$L$_{\odot}$), as done in Magnelli et
al. (2013) for the total galaxy
population. We point out that in the case of the ``faint'' galaxy
population, the comoving densities mainly rely on the extrapolation
of the LF, in particular at
$z > 0.4$. Thus, we recommend caution when interpreting values not
directly constrained by the {\it{Spitzer}} and {\it{Herschel}}
observations. As already pointed out in Magnelli et al. (2013), we
emphasize that the LIRG and ULIRG designations are used here strictly
to segregate the luminosity bins, but not to imply physical
properties. Indeed, \textit{Herschel} studies have unambiguously
revealed that high-redshift (U)LIRGs do not have the same properties
as their local counterparts
\citep[e.g.][]{Elbaz+11,wuyts+11}.

We find that the number density of ULIRGs and LIRGs evolves strongly
from redshift $\sim 1,$ and the evolution is faster for groups than
for the total galaxy population.  The number density evolution is
faster for IR-brighter galaxies, both in the field and in groups. We
fit the number density vs. redshift of any given subsample with a
power law of the type $\rm{number density} \propto
(1+z)^{\alpha}$, up to $z \sim 1$.  Beyond $z \sim 1,$ the number
densities of the different IR galaxy populations do not evolve any
further. We find $\alpha=0.73\pm0.08, 5.4\pm0.5,$ and $8.9\pm0.7$ for
the faint IR galaxies, LIRGs, and ULIRGs, respectively, in the field,
and $\alpha=4\pm1, 12\pm2, 22\pm4$ for the faint IR galaxies, LIRGs,
and ULIRGs, respectively, in groups.  The best-fit $\alpha$ values are
significantly higher for group galaxies than for the total galaxy
population, confirming the much faster decline of the number density
of IR emitting galaxies in groups relative to the total population. The comoving
number densities of faint IR galaxies, LIRGs, and ULIRGs decrease by
factors $\sim$ 1.5, 34 and 215, respectively, since $z \sim 1$, in the
field, and by a factor 54, and 3.5 and 6 orders of magnitudes,
respectively, in groups. 

At $z\sim 1$, group galaxies contribute 40\% to 60\% of the whole IR galaxy
population in the sub-ULIRG regime, but almost all ULIRGs are in
groups. This is consistent with a reversal (Elbaz et al. 2007) or a
flattening (Ziparo et al. 2014) of the SFR-density relation. In
addition, there is also consistency with our previous findings that
the fraction of bright IR galaxies ($L_{IR} > 10^{11} L_{\odot}$) is
higher in a denser environment at $z \sim 1$ \citep{Popesso+11}. That ULIRGs are primarily located in massive haloes at high
redshift is also consistent with the recent findings of \citet{Magliocchetti+13,Magliocchetti+14}, who find that the clustering lengths of star-forming systems present a sharp increase as a function of
redshift. This behaviour is reflected in the trend of the masses of
the dark matter hosts of star-forming galaxies, which increase from
$10^{11-11.5} M_{\odot}$ at $ z < 1$ to $\sim 10^{13.5} M_{\odot}$
between $z \sim 1$ and $z \sim 2$. Our analysis shows that galaxies
which actively form stars at high redshifts are not the same
population of sources we observe in the more local universe. In fact,
vigorous star formation in the early universe is hosted by very
massive structures, while for $z < 1$ a comparable activity is found
in much smaller systems, consistent with the downsizing scenario.

\begin{figure}
\begin{center}
\includegraphics[width=0.49\textwidth]{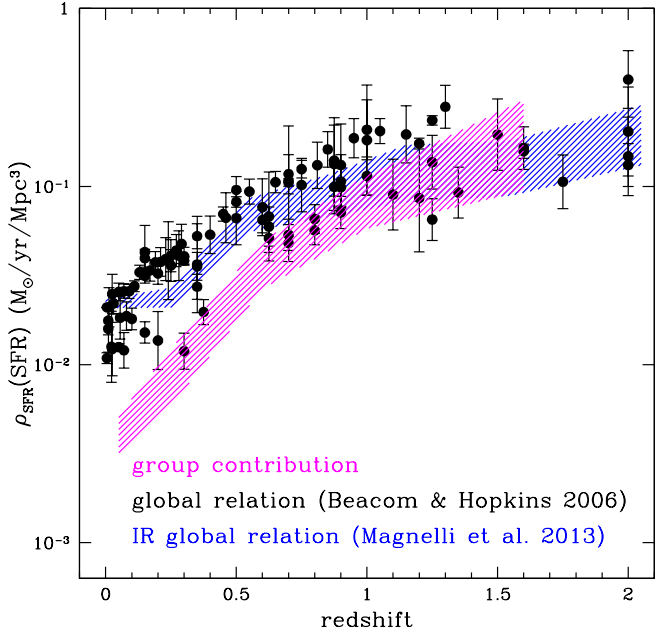}
\caption{Contribution of the group galaxy population (magenta shaded
  region) to the cosmic SFR density estimated by
  Magnelli et al. (2013, blue shaded region, obscured SFR density) and 
  Beacom \& Hopkins (2006, black points, total SFR density).}
\label{lum_density}
\end{center}
\end{figure}

By integrating our group galaxy IR LFs, we derive the evolution of the
comoving IR luminosity density.  We use the relation of
\citet{Kennicutt98} to convert $L_{IR}$ into SFR and derive the
contribution of the group galaxy population to the cosmic SFR density
of the Universe in redshit bins.  We assume that the IR flux is
completely dominated by obscured SF rather than by AGN activity, also
for the 5\% of group members identified as X-ray-emitting AGNs.
The hosts of 87\% of these AGNs have been detected by PACS,
and it has been shown that their IR emission originates in the host
galaxy in $>\,$94\% of the cases, that is it has a SF origin \citep{Shao+10,mullaney+12,rosario+12}. For the
remaining 13\% of AGNs, the 24 ${\mu}$m flux detected with
{\it{Spitzer}} could in principle be contaminated by the AGN
emission. However, since these galaxies are faint IR sources, and they
represent only 0.65\% of the entire group galaxy population studied in
this work, their contribution is too marginal to have any significant
effect on our estimates of the SFR density contributed by groups.

To properly derive the total SFR density of the Universe,
Magnelli et al. (2013) combine the obscured SFR density with the
unobscured SFR density of the Universe derived by
\citet{Cucciati+12} using rest-frame UV observations. We do not have
an estimate of the unobscured SFR density for the population of group
galaxies.  Thus, for a fair comparison, we consider the contribution
of the group galaxy population only to the obscured cosmic SFR
density. This does, however constitute most (75--88\%) of the full cosmic
SFR density at any redshift (Magnelli et al. 2013).

 In Fig. \ref{lum_density} we show the contribution of the group galaxy
population to the obscured cosmic SFR density. For completeness we also show the
compilation of the SFR density estimates of Beacom \& Hopkins (2006), which
are based on both IR and UV data. At $z\gtrsim 1$ the group galaxy
population makes a substantial (60-80\%)
contribution to the cosmic SFR density, because
most of the $z > 1$ cosmic SFR density is provided by (U)LIRGs
(Magnelli et al. 2009, 2011, 2013, Gruppioni et al. 2010, 2013), and a
substantial fraction (70\%) of LIRGs and the
totality of ULIRGS of $z > 1$ are located in groups (see
Fig. \ref{num_density}). At $z\lesssim 1$ the rapid decline in the
number density of (U)LIRGs in groups drives the similarly rapid
decline in the group SFR density.  While the cosmic SFR density
declines by $\sim 0.65$ dex since $z=1$, the contribution of group
galaxies to the cosmic SFR density decreases by $1.43$ dex, and
becomes negligible by $z \sim 0$.

\section{Discussion}
\label{s:discussion}


At $z< 0.4,$ the group IR LF is characterized by a much
steeper cut-off at the bright end than the total IR LF. In other
words, the very bright, rare, and strongly star-forming IR-emitting
galaxies that are observed in the field do not reside in groups at $z<
0.4$. The volume density of group IR emitting galaxies is $\le 10\%$
that of the total population. At $0.4 < z < 0.8,$ we find consistency
between the shapes (characterized by $L^*$ and faint end slope) of the
group IR LF and the total IR LF, and the volume density of group IR
emitting galaxies is only $\sim 60$\% lower than for the total
population.  At $z \sim 1$ the galaxy group IR LF and the Kurk et
al. (2008) structure LF exhibit a slighty brighter $L^*$ luminosity
than the total LFs of \citet{gruppioni+13}, and \citet{magnelli+13},
respectively. This suggests that star-forming group galaxies have a
slightly higher mean SFR than star-forming galaxies in the field.  The
volume density of IR-emitting group galaxies is very close to that of
the total galaxy population; that is to say,  the group galaxy population
provides a $\sim$ 70\% of the LIRGs and the totality of the ULIRGS of the total IR-emitting
galaxy population.

The comoving number density of (U)LIRGs in groups evolves
strongly, since redshift $\sim 1,$ and this evolution is much faster
than observed for the total population of (U)LIRGs. The
evolution is faster for IR-brighter galaxies, both in groups and in
the field.  Group LIRGs and other, less bright, group IR-emitting
galaxies account for 40\% to 60\% of the whole IR galaxy population at
$z\sim 1$, but almost all ULIRGs at the same redshift are located in
groups.  This is consistent with our previous result of a higher
fraction of LIRGs in denser environments at $z\sim 1$ (Popesso et
al. 2011).  This is further evidence that the mean SFR in groups is
higher than in the field at $z\sim 1$, consistent with previous
findings that have suggested a reversal (Elbaz et al. 2007) or flattening
(Ziparo et al. 2014) of the SFR-density relation.  In addition, since most of high-redshift LIRGs are MS galaxies and are generally
not associated to merger activity, our findings would suggest
that galaxy groups are not favorite sites for the onset of a strong
merger activity, at least not among gas-rich galaxies, and not at
$z<1$.

We show that at $z\gtrsim 1$ group galaxies contribute 60\% to 80\% to the cosmic SFR density, because
(U)LIRGs are the main contributors to the cosmic SFR density at $z >
1$ \citep{magnelli+13,gruppioni+13},
and according to our results, they are largely group galaxies at that
epoch.  At $z\lesssim 1$ the number density of (U)LIRGs in groups
declines very rapidly, and this then leads to a similarly rapid
decline in the contribution of group galaxies to the cosmic SFR
density.  By $z \sim 0$ this contribution becomes entirely negligible.

Our results agree with previous claims about a reversal or
a flattening of the SFR-density relation
\citep{Elbaz+07,Popesso+11,ziparo+13} at z$\sim$1. Based on the same
data set as used in this paper, \citet{ziparo+13} conclude that the
differential evolution of groups galaxies with respect to field
galaxies is due to a faster quenching of main-sequence star-forming
galaxies in groups than in the field. (High-$z$ (U)LIRGs are
main-sequence galaxies.) 

These and our results clearly indicate that the SF activity of
galaxies evolves differently in different environments. This is also
supported by independent, observational evidence.; for example, the galaxy
red sequence forms earlier in groups than in the field, especially at
high stellar masses \citep{iovino+10,kovac+10}, and there is a
transient population of ``red spirals'' in groups not observed in the
field \citep{Balogh+11,wolf+09,mei+12}, which suggests morphological
transformations are in place in groups at least after $z \sim
1$. Quenching of SF activity therefore occurs earlier in galaxies
that are embedded in more massive haloes than on average. In this sense, as in
Popesso et al. (2012), we see evidence of a ``halo downsizing''
effect, whereby massive haloes evolve more rapidly than haloes with lower
masses \citep{NvdBD06}.  This ``halo downsizing'' effect is not at
odds with the current hierarchical paradigm of structure formation. It
implies that the quenching process is driven by the accretion of
galaxies from the cosmic web into more massive haloes.  Halo downsizing
then comes naturally in the hierarchical scenario. In fact, massive
galaxies are hosted by massive haloes, and haloes in overdense regions
on average form earlier and merge more rapidly than haloes in regions
of average density \citep{Gao+04}.

To isolate the physical processes responsible for the quenching of SF
activity, Peng et al. (2010) have identified two types of quenching,
one dependent on the galaxy ``mass'' and another dependent on the
galaxy ``environment''.  The AGN feedback can be ascribed to the
``mass'' quenching type. Powerful jets (radio mode) or outflows
(quasar mode) driven by the AGN activity swipe out the gas from the AGN
host galaxy, at the same time halting gas accretion onto the central
black hole and the SF activity of the host galaxy. This process only
depends on the individual galaxy properties, such as its stellar mass,
which is proportional to the mass of the central black hole. This
process cannot be responsible for the ``halo downsizing'' we
observe. The faster evolution of the group galaxy population is
observed for galaxies of any IR luminosity, hence, of any stellar
mass, given the relation known as the main sequence of SF galaxies (Noeske
et al. 2007, Elbaz et al. 2007), which is also obeyed (at high-$z$) by
(U)LIRGs. If AGN feedback is only dependent on galaxy mass, the
evolution should be similar among (U)LIRGs in the field and in groups,
but it is not.

We therefore need to appeal to ``environment'' quenching to explain
our results. Related to the ``environment'' quenching are the
processes identified by the cold-hot two-mode gas accretion model
\citep{keres+05,db+06}. According to this model, large haloes primarily
accrete hot gas while small haloes primarily accrete cold gas.  In the
cold accretion paradigm, the gas cooling along filaments tends to fall
to the centre of the halo, so that galaxies that turn into
satellites become disconnected from their feeding filaments.

In semi-analytical models, such as those based on the Millennium
Simulation \citep{springel+05}, this physical process is
implemented by a sudden cut-off of the cold gas supply as a galaxy
enters massive haloes dominated by hot accretion. This leads to an
immediate quenching of gas accretion, and, consequently, of SF
activity. This implementation
of the ``environment'' quenching leads to a rapid evolution of the
galaxy population of massive haloes, which turn into a 90\% red and dead
galaxy population already by redshift $\sim 2$. As a consequence,
there is an excess of red and passive galaxies in groups and clusters in the
simulated local Universe with respect to the observations
\citep{wang+07}, the so-called ``satellite over-quenching'' problem.
Therefore, in these models, 
galaxies in haloes with mass higher than $10^{12}$ $M_{\odot}$ provide
a very marginal contribution to the CSFR density at any redshift \citep{voort+11}, and this is clearly at odds with our results. 

Recently, using SPH simulations, \citet{simha+09} have shown that
satellite galaxies continue to accrete gas and convert it into stars
for quite some time (0.5-1 Gyr) after entering a larger halo, where
gas is hot. The gas accretion declines steadily over this period,
leading to gradual quenching.  Observational support for a longer
quenching timescale than predicted by traditional semi-analytical
models has recently come from the analysis \citet{wetzel+13}. They do
not suggest a gradual quenching, but a rapid one ($< 0.8$ Gyr), which
however occurs only $2-4$ Gyr after the satellite infall into
groups/clusters.  As argued by \citet{simha+09}, allowing for a longer
quenching timescale after satellite accretion should improve the
agreement of semi-analytic models with the observed colour
distributions of satellite galaxies in groups and with the observed
colour dependence of galaxy clustering. Better agreement with the
observed evolution of the SF activity in groups is also to be
expected.

\section{Summary and conclusions}
\label{s:conclusion}
We have determined the IR LF of group galaxies in the redshift range
$z=0$--1.6, based on a sample of 39 X-ray selected groups with
extensive spectroscopic and photometric data available, in particular,
from {\it Spitzer} MIPS and {\it Herschel} PACS.

We find a differential evolution of the IR LFs of group and field
galaxies. The group IR LF at low-$z$ lacks very IR-bright galaxies in the field, and group IR-emitting galaxies contribute
$\lesssim 10$\% of the comoving volume density of the IR galaxy
population as a whole. The fraction of very IR-bright galaxies (LIRGs
and ULIRGs) increases rapidly with $z$, and this increase is much
faster in groups than in the general population. As a result, the
shape of the group galaxy IR LF first becomes similar to that of field
galaxies (at intermediate redshifts, $0.4 < z <0.8$) and then, at $z
\sim 1$, it shows an excess of very IR-bright galaxies with respect to
the field IR LF. The contribution of group IR-emitting galaxies to the
comoving volume density of the IR galaxy population as a whole also
increases with redshift, reaching $\sim$ 40\% at $z\sim 0.5$ and $\sim$ 60-70\% at $z\sim 1$.

We quantify this differential evolution of the group and field IR LFs
also in terms of the comoving number density of (U)LIRGs in groups and
in the general population. We find that almost all ULIRGs are located
in groups at $z \sim 1$, while they are almost all outside groups at
$z \sim 0$.
Finally, we quantify the contribution of group galaxies to the
cosmic SFR density, and find it to increasing from virtually none
at $z \sim 0$ to 60-80\%at $z \sim 1$.

Our results indicate ``halo downsizing'' in the star
formation processes of galaxies.  Since the differential evolution we
observe not only concerns very bright (hence massive) galaxies
(i.e. LIRGs and ULIRGs) but also IR-emitting galaxies of lower
luminosities, we argue that ``mass'' quenching alone cannot explain
our results, so ``environment'' quenching is also required.

\begin{acknowledgements}
The authors aknowledge G. Zamorani for the very useful comments on the
early draft.  PACS has been developed by a consortium of institutes
led by MPE (Germany) and including UVIE (Austria); KUL, CSL, IMEC
(Belgium); CEA, OAMP (France); MPIA (Germany); IFSI, OAP/AOT,
OAA/CAISMI, LENS, SISSA (Italy); IAC (Spain). This development has
been supported by the funding agencies BMVIT (Austria), ESA-PRODEX
(Belgium), CEA/CNES (France), DLR (Germany), ASI (Italy), and
CICYT/MCYT (Spain).

We gratefully acknowledge the contributions of the entire COSMOS
collaboration consisting of more than 100 scientists. More information
about the COSMOS survey is available at
http://www.astro.caltech.edu/$\sim$cosmos.

This research made use of NASA's Astrophysics Data System, of NED,
which is operated by JPL/Caltech, under contract with NASA, and of
SDSS, which has been funded by the Sloan Foundation, NSF, the US
Department of Energy, NASA, the Japanese Monbukagakusho, the Max
Planck Society, and the Higher Education Funding Council of England.
The SDSS is managed by the participating institutions
(www.sdss.org/collaboration/credits.html).

\end{acknowledgements}

\bibliography{master}

\begin{thebibliography}{101}
\expandafter\ifx\csname natexlab\endcsname\relax\def\natexlab#1{#1}\fi

\bibitem[{{Bai} {et~al.}(2009){Bai}, {Rieke}, {Rieke}, {Christlein}, \&
  {Zabludoff}}]{Bai+09}
{Bai}, L., {Rieke}, G.~H., {Rieke}, M.~J., {Christlein}, D., \& {Zabludoff},
  A.~I. 2009, \apj, 693, 1840

\bibitem[{{Bai} {et~al.}(2006){Bai}, {Rieke}, {Rieke}, {Hinz}, {Kelly}, \&
  {Blaylock}}]{Bai+06}
{Bai}, L., {Rieke}, G.~H., {Rieke}, M.~J., {et~al.} 2006, \apj, 639, 827

\bibitem[{{Balestra} {et~al.}(2010){Balestra}, {Mainieri}, {Popesso},
  {Dickinson}, {Nonino}, {Rosati}, {Teimoorinia}, {Vanzella}, {Cristiani},
  {Cesarsky}, {Fosbury}, {Kuntschner}, \& {Rettura}}]{balestra+10}
{Balestra}, I., {Mainieri}, V., {Popesso}, P., {et~al.} 2010, \aap, 512, A12

\bibitem[{{Balogh} {et~al.}(2011){Balogh}, {McGee}, {Wilman}, {Finoguenov},
  {Parker}, {Connelly}, {Mulchaey}, {Bower}, {Tanaka}, \&
  {Giodini}}]{Balogh+11}
{Balogh}, M.~L., {McGee}, S.~L., {Wilman}, D.~J., {et~al.} 2011, \mnras, 412,
  2303

\bibitem[{{Barger} {et~al.}(2008){Barger}, {Cowie}, \& {Wang}}]{barger+08}
{Barger}, A.~J., {Cowie}, L.~L., \& {Wang}, W.-H. 2008, \apj, 689, 687

\bibitem[{{Bell}(2003)}]{bell+03}
{Bell}, E.~F. 2003, \apj, 586, 794

\bibitem[{{Bell} {et~al.}(2005){Bell}, {Papovich}, {Wolf}, {Le Floc'h},
  {Caldwell}, {Barden}, {Egami}, {McIntosh}, {Meisenheimer},
  {P{\'e}rez-Gonz{\'a}lez}, {Rieke}, {Rieke}, {Rigby}, \& {Rix}}]{bell+05}
{Bell}, E.~F., {Papovich}, C., {Wolf}, C., {et~al.} 2005, \apj, 625, 23

\bibitem[{{Berta} {et~al.}(2010){Berta}, {Magnelli}, {Lutz}, {Altieri},
  {Aussel}, {Andreani}, {Bauer}, {Bongiovanni}, {Cava}, {Cepa}, {Cimatti},
  {Daddi}, {Dominguez}, {Elbaz}, {Feuchtgruber}, {F{\"o}rster Schreiber},
  {Genzel}, {Gruppioni}, {Katterloher}, {Magdis}, {Maiolino}, {Nordon},
  {P{\'e}rez Garc{\'{\i}}a}, {Poglitsch}, {Popesso}, {Pozzi}, {Riguccini},
  {Rodighiero}, {Saintonge}, {Santini}, {Sanchez-Portal}, {Shao}, {Sturm},
  {Tacconi}, {Valtchanov}, {Wetzstein}, \& {Wieprecht}}]{Berta+10}
{Berta}, S., {Magnelli}, B., {Lutz}, D., {et~al.} 2010, \aap, 518, L30

\bibitem[{{Biviano} {et~al.}(2011){Biviano}, {Fadda}, {Durret}, {Edwards}, \&
  {Marleau}}]{Biviano+11}
{Biviano}, A., {Fadda}, D., {Durret}, F., {Edwards}, L.~O.~V., \& {Marleau}, F.
  2011, \aap, 532, A77

\bibitem[{{Buat} {et~al.}(2002){Buat}, {Boselli}, {Gavazzi}, \&
  {Bonfanti}}]{buat+02}
{Buat}, V., {Boselli}, A., {Gavazzi}, G., \& {Bonfanti}, C. 2002, \aap, 383,
  801

\bibitem[{{Capak} {et~al.}(2007){Capak}, {Aussel}, {Ajiki}, {McCracken},
  {Mobasher}, {Scoville}, {Shopbell}, {Taniguchi}, {Thompson}, {Tribiano},
  {Sasaki}, {Blain}, {Brusa}, {Carilli}, {Comastri}, {Carollo}, {Cassata},
  {Colbert}, {Ellis}, {Elvis}, {Giavalisco}, {Green}, {Guzzo}, {Hasinger},
  {Ilbert}, {Impey}, {Jahnke}, {Kartaltepe}, {Kneib}, {Koda}, {Koekemoer},
  {Komiyama}, {Leauthaud}, {Le Fevre}, {Lilly}, {Liu}, {Massey}, {Miyazaki},
  {Murayama}, {Nagao}, {Peacock}, {Pickles}, {Porciani}, {Renzini}, {Rhodes},
  {Rich}, {Salvato}, {Sanders}, {Scarlata}, {Schiminovich}, {Schinnerer},
  {Scodeggio}, {Sheth}, {Shioya}, {Tasca}, {Taylor}, {Yan}, \&
  {Zamorani}}]{Capak+07}
{Capak}, P., {Aussel}, H., {Ajiki}, M., {et~al.} 2007, \apjs, 172, 99

\bibitem[{{Caputi} {et~al.}(2007){Caputi}, {Lagache}, {Yan}, {Dole},
  {Bavouzet}, {Le Floc'h}, {Choi}, {Helou}, \& {Reddy}}]{Caputi+07}
{Caputi}, K.~I., {Lagache}, G., {Yan}, L., {et~al.} 2007, \apj, 660, 97

\bibitem[{{Cardamone} {et~al.}(2010){Cardamone}, {van Dokkum}, {Urry},
  {Taniguchi}, {Gawiser}, {Brammer}, {Taylor}, {Damen}, {Treister}, {Cobb},
  {Bond}, {Schawinski}, {Lira}, {Murayama}, {Saito}, \&
  {Sumikawa}}]{cardamone+10}
{Cardamone}, C.~N., {van Dokkum}, P.~G., {Urry}, C.~M., {et~al.} 2010, \apjs,
  189, 270

\bibitem[{{Chung} {et~al.}(2010){Chung}, {Gonzalez}, {Clowe}, {Markevitch}, \&
  {Zaritsky}}]{Chung+10}
{Chung}, S.~M., {Gonzalez}, A.~H., {Clowe}, D., {Markevitch}, M., \&
  {Zaritsky}, D. 2010, \apj, 725, 1536

\bibitem[{{Cimatti} {et~al.}(2008){Cimatti}, {Robberto}, {Baugh}, {Beckwith},
  {Content}, {Daddi}, {De Lucia}, {Garilli}, {Guzzo}, {Kauffmann}, {Lehnert},
  {Maccagni}, {Mart{\'{\i}}nez-Sansigre}, {Pasian}, {Reid}, {Rosati},
  {Salvaterra}, {Stiavelli}, {Wang}, {Osorio}, {Balcells}, {Bersanelli},
  {Bertoldi}, {Blaizot}, {Bottini}, {Bower}, {Bulgarelli}, {Burgasser},
  {Burigana}, {Butler}, {Casertano}, {Ciardi}, {Cirasuolo}, {Clampin}, {Cole},
  {Comastri}, {Cristiani}, {Cuby}, {Cuttaia}, {de Rosa}, {Sanchez}, {di Capua},
  {Dunlop}, {Fan}, {Ferrara}, {Finelli}, {Franceschini}, {Franx}, {Franzetti},
  {Frenk}, {Gardner}, {Gianotti}, {Grange}, {Gruppioni}, {Gruppuso}, {Hammer},
  {Hillenbrand}, {Jacobsen}, {Jarvis}, {Kennicutt}, {Kimble}, {Kriek}, {Kurk},
  {Kneib}, {Le Fevre}, {Macchetto}, {MacKenty}, {Madau}, {Magliocchetti},
  {Maino}, {Mandolesi}, {Masetti}, {McLure}, {Mennella}, {Meyer}, {Mignoli},
  {Mobasher}, {Molinari}, {Morgante}, {Morris}, {Nicastro}, {Oliva},
  {Padovani}, {Palazzi}, {Paresce}, {Garrido}, {Pian}, {Popa}, {Postman},
  {Pozzetti}, {Rayner}, {Rebolo}, {Renzini}, {R{\"o}ttgering}, {Schinnerer},
  {Scodeggio}, {Saisse}, {Shanks}, {Shapley}, {Sharples}, {Shea}, {Silk},
  {Smail}, {Span{\'o}}, {Steinacker}, {Stringhetti}, {Szalay}, {Tresse},
  {Trifoglio}, {Urry}, {Valenziano}, {Villa}, {Perez}, {Walter}, {Ward},
  {White}, {White}, {Wright}, {Wyse}, {Zamorani}, {Zacchei}, {Zeilinger}, \&
  {Zerbi}}]{Cimatti+08}
{Cimatti}, A., {Robberto}, M., {Baugh}, C., {et~al.} 2008, Experimental
  Astronomy, 37

\bibitem[{{Clements} {et~al.}(1996){Clements}, {Sutherland}, {McMahon}, \&
  {Saunders}}]{clements+96}
{Clements}, D.~L., {Sutherland}, W.~J., {McMahon}, R.~G., \& {Saunders}, W.
  1996, \mnras, 279, 477

\bibitem[{{Colless}(1989)}]{colless+89}
{Colless}, M. 1989, \mnras, 237, 799

\bibitem[{{Cooper} {et~al.}(2008){Cooper}, {Newman}, {Weiner}, {Yan},
  {Willmer}, {Bundy}, {Coil}, {Conselice}, {Davis}, {Faber}, {Gerke},
  {Guhathakurta}, {Koo}, \& {Noeske}}]{Cooper+08}
{Cooper}, M.~C., {Newman}, J.~A., {Weiner}, B.~J., {et~al.} 2008, \mnras, 383,
  1058

\bibitem[{{Cooper} {et~al.}(2012){Cooper}, {Yan}, {Dickinson}, {Juneau},
  {Lotz}, {Newman}, {Papovich}, {Salim}, {Walth}, {Weiner}, \&
  {Willmer}}]{cooper+12}
{Cooper}, M.~C., {Yan}, R., {Dickinson}, M., {et~al.} 2012, \mnras, 425, 2116

\bibitem[{{Cowie} {et~al.}(2004){Cowie}, {Barger}, {Fomalont}, \&
  {Capak}}]{cowie+04}
{Cowie}, L.~L., {Barger}, A.~J., {Fomalont}, E.~B., \& {Capak}, P. 2004, \apjl,
  603, L69

\bibitem[{{Cucciati} {et~al.}(2012){Cucciati}, {Tresse}, {Ilbert}, {Le
  F{\`e}vre}, {Garilli}, {Le Brun}, {Cassata}, {Franzetti}, {Maccagni},
  {Scodeggio}, {Zucca}, {Zamorani}, {Bardelli}, {Bolzonella}, {Bielby},
  {McCracken}, {Zanichelli}, \& {Vergani}}]{Cucciati+12}
{Cucciati}, O., {Tresse}, L., {Ilbert}, O., {et~al.} 2012, \aap, 539, A31

\bibitem[{{De Propris} {et~al.}(2003){De Propris}, {Colless}, {Driver},
  {Couch}, {Peacock}, {Baldry}, {Baugh}, {Bland-Hawthorn}, {Bridges}, {Cannon},
  {Cole}, {Collins}, {Cross}, {Dalton}, {Efstathiou}, {Ellis}, {Frenk},
  {Glazebrook}, {Hawkins}, {Jackson}, {Lahav}, {Lewis}, {Lumsden}, {Maddox},
  {Madgwick}, {Norberg}, {Percival}, {Peterson}, {Sutherland}, \&
  {Taylor}}]{depropris+03}
{De Propris}, R., {Colless}, M., {Driver}, S.~P., {et~al.} 2003, \mnras, 342,
  725

\bibitem[{{Dekel} \& {Birnboim}(2006)}]{db+06}
{Dekel}, A. \& {Birnboim}, Y. 2006, \mnras, 368, 2

\bibitem[{{Dressler}(1980)}]{Dressler80}
{Dressler}, A. 1980, \apj, 236, 351

\bibitem[{{Eke} {et~al.}(2005){Eke}, {Baugh}, {Cole}, {Frenk}, {King}, \&
  {Peacock}}]{eke+05}
{Eke}, V.~R., {Baugh}, C.~M., {Cole}, S., {et~al.} 2005, \mnras, 362, 1233

\bibitem[{{Elbaz} {et~al.}(2007){Elbaz}, {Daddi}, {Le Borgne}, {Dickinson},
  {Alexander}, {Chary}, {Starck}, {Brandt}, {Kitzbichler}, {MacDonald},
  {Nonino}, {Popesso}, {Stern}, \& {Vanzella}}]{Elbaz+07}
{Elbaz}, D., {Daddi}, E., {Le Borgne}, D., {et~al.} 2007, \aap, 468, 33

\bibitem[{{Elbaz} {et~al.}(2011){Elbaz}, {Dickinson}, {Hwang},
  {D{\'{\i}}az-Santos}, {Magdis}, {Magnelli}, {Le Borgne}, {Galliano},
  {Pannella}, {Chanial}, {Armus}, {Charmandaris}, {Daddi}, {Aussel}, {Popesso},
  {Kartaltepe}, {Altieri}, {Valtchanov}, {Coia}, {Dannerbauer}, {Dasyra},
  {Leiton}, {Mazzarella}, {Alexander}, {Buat}, {Burgarella}, {Chary}, {Gilli},
  {Ivison}, {Juneau}, {Le Floc'h}, {Lutz}, {Morrison}, {Mullaney}, {Murphy},
  {Pope}, {Scott}, {Brodwin}, {Calzetti}, {Cesarsky}, {Charlot}, {Dole},
  {Eisenhardt}, {Ferguson}, {F{\"o}rster Schreiber}, {Frayer}, {Giavalisco},
  {Huynh}, {Koekemoer}, {Papovich}, {Reddy}, {Surace}, {Teplitz}, {Yun}, \&
  {Wilson}}]{Elbaz+11}
{Elbaz}, D., {Dickinson}, M., {Hwang}, H.~S., {et~al.} 2011, \aap, 533, A119

\bibitem[{{Farrah} {et~al.}(2001){Farrah}, {Rowan-Robinson}, {Oliver},
  {Serjeant}, {Borne}, {Lawrence}, {Lucas}, {Bushouse}, \&
  {Colina}}]{farrah+01}
{Farrah}, D., {Rowan-Robinson}, M., {Oliver}, S., {et~al.} 2001, \mnras, 326,
  1333

\bibitem[{{Feruglio} {et~al.}(2010){Feruglio}, {Aussel}, {Le Floc'h}, {Ilbert},
  {Salvato}, {Capak}, {Fiore}, {Kartaltepe}, {Sanders}, {Scoville},
  {Koekemoer}, \& {Ideue}}]{Feruglio+10}
{Feruglio}, C., {Aussel}, H., {Le Floc'h}, E., {et~al.} 2010, \apj, 721, 607

\bibitem[{{Finn} {et~al.}(2010){Finn}, {Desai}, {Rudnick}, {Poggianti}, {Bell},
  {Hinz}, {Jablonka}, {Milvang-Jensen}, {Moustakas}, {Rines}, \&
  {Zaritsky}}]{Finn+10}
{Finn}, R.~A., {Desai}, V., {Rudnick}, G., {et~al.} 2010, \apj, 720, 87

\bibitem[{{Finoguenov} {et~al.}(2009){Finoguenov}, {Connelly}, {Parker},
  {Wilman}, {Mulchaey}, {Saglia}, {Balogh}, {Bower}, \&
  {McGee}}]{finoguenov+09}
{Finoguenov}, A., {Connelly}, J.~L., {Parker}, L.~C., {et~al.} 2009, \apj, 704,
  564

\bibitem[{{Finoguenov} {et~al.}(2010){Finoguenov}, {Watson}, {Tanaka},
  {Simpson}, {Cirasuolo}, {Dunlop}, {Peacock}, {Farrah}, {Akiyama}, {Ueda},
  {Smol{\v c}i{\'c}}, {Stewart}, {Rawlings}, {van Breukelen}, {Almaini},
  {Clewley}, {Bonfield}, {Jarvis}, {Barr}, {Foucaud}, {McLure}, {Sekiguchi}, \&
  {Egami}}]{finoguenov+10}
{Finoguenov}, A., {Watson}, M.~G., {Tanaka}, M., {et~al.} 2010, \mnras, 403,
  2063

\bibitem[{{Gao} {et~al.}(2004){Gao}, {De Lucia}, {White}, \&
  {Jenkins}}]{Gao+04}
{Gao}, L., {De Lucia}, G., {White}, S.~D.~M., \& {Jenkins}, A. 2004, \mnras,
  352, L1

\bibitem[{{G{\'o}mez} {et~al.}(2003){G{\'o}mez}, {Nichol}, {Miller}, {Balogh},
  {Goto}, {Zabludoff}, {Romer}, {Bernardi}, {Sheth}, {Hopkins}, {Castander},
  {Connolly}, {Schneider}, {Brinkmann}, {Lamb}, {SubbaRao}, \&
  {York}}]{Gomez+03}
{G{\'o}mez}, P.~L., {Nichol}, R.~C., {Miller}, C.~J., {et~al.} 2003, \apj, 584,
  210

\bibitem[{{Gruppioni} {et~al.}(2013){Gruppioni}, {Pozzi}, {Rodighiero},
  {Delvecchio}, {Berta}, {Pozzetti}, {Zamorani}, {Andreani}, {Cimatti},
  {Ilbert}, {Le Floc'h}, {Lutz}, {Magnelli}, {Marchetti}, {Monaco}, {Nordon},
  {Oliver}, {Popesso}, {Riguccini}, {Roseboom}, {Rosario}, {Sargent},
  {Vaccari}, {Altieri}, {Aussel}, {Bongiovanni}, {Cepa}, {Daddi},
  {Dom{\'{\i}}nguez-S{\'a}nchez}, {Elbaz}, {F{\"o}rster Schreiber}, {Genzel},
  {Iribarrem}, {Magliocchetti}, {Maiolino}, {Poglitsch}, {P{\'e}rez
  Garc{\'{\i}}a}, {Sanchez-Portal}, {Sturm}, {Tacconi}, {Valtchanov},
  {Amblard}, {Arumugam}, {Bethermin}, {Bock}, {Boselli}, {Buat}, {Burgarella},
  {Castro-Rodr{\'{\i}}guez}, {Cava}, {Chanial}, {Clements}, {Conley}, {Cooray},
  {Dowell}, {Dwek}, {Eales}, {Franceschini}, {Glenn}, {Griffin},
  {Hatziminaoglou}, {Ibar}, {Isaak}, {Ivison}, {Lagache}, {Levenson}, {Lu},
  {Madden}, {Maffei}, {Mainetti}, {Nguyen}, {O'Halloran}, {Page}, {Panuzzo},
  {Papageorgiou}, {Pearson}, {P{\'e}rez-Fournon}, {Pohlen}, {Rigopoulou},
  {Rowan-Robinson}, {Schulz}, {Scott}, {Seymour}, {Shupe}, {Smith}, {Stevens},
  {Symeonidis}, {Trichas}, {Tugwell}, {Vigroux}, {Wang}, {Wright}, {Xu},
  {Zemcov}, {Bardelli}, {Carollo}, {Contini}, {Le F{\'e}vre}, {Lilly},
  {Mainieri}, {Renzini}, {Scodeggio}, \& {Zucca}}]{gruppioni+13}
{Gruppioni}, C., {Pozzi}, F., {Rodighiero}, G., {et~al.} 2013, \mnras, 432, 23

\bibitem[{{Gruppioni} {et~al.}(2011){Gruppioni}, {Pozzi}, {Zamorani}, \&
  {Vignali}}]{Gruppioni+11}
{Gruppioni}, C., {Pozzi}, F., {Zamorani}, G., \& {Vignali}, C. 2011, \mnras,
  416, 70

\bibitem[{{Guo} {et~al.}(2014){Guo}, {Lacey}, {Norberg}, {Cole}, {Baugh},
  {Frenk}, {Cooray}, {Dye}, {Bourne}, {Dunne}, {Eales}, {Ivison}, {Maddox},
  {Alpasan}, {Baldry}, {Driver}, \& {Robotham}}]{Guo+14}
{Guo}, Q., {Lacey}, C., {Norberg}, P., {et~al.} 2014, ArXiv e-prints

\bibitem[{{Haines} {et~al.}(2013){Haines}, {Pereira}, {Smith}, {Egami},
  {Sanderson}, {Babul}, {Finoguenov}, {Merluzzi}, {Busarello}, {Rawle}, \&
  {Okabe}}]{Haines+13}
{Haines}, C.~P., {Pereira}, M.~J., {Smith}, G.~P., {et~al.} 2013, \apj, 775,
  126

\bibitem[{{Haines} {et~al.}(2010){Haines}, {Smith}, {Pereira}, {Egami},
  {Moran}, {Hardegree-Ullman}, {Rawle}, \& {Rex}}]{Haines+10}
{Haines}, C.~P., {Smith}, G.~P., {Pereira}, M.~J., {et~al.} 2010, \aap, 518,
  L19

\bibitem[{{Heavens} {et~al.}(2004){Heavens}, {Panter}, {Jimenez}, \&
  {Dunlop}}]{heavens+04}
{Heavens}, A., {Panter}, B., {Jimenez}, R., \& {Dunlop}, J. 2004, \nat, 428,
  625

\bibitem[{{Hinshaw} {et~al.}(2013){Hinshaw}, {Larson}, {Komatsu}, {Spergel},
  {Bennett}, {Dunkley}, {Nolta}, {Halpern}, {Hill}, {Odegard}, {Page}, {Smith},
  {Weiland}, {Gold}, {Jarosik}, {Kogut}, {Limon}, {Meyer}, {Tucker}, {Wollack},
  \& {Wright}}]{Hinshaw+13}
{Hinshaw}, G., {Larson}, D., {Komatsu}, E., {et~al.} 2013, \apjs, 208, 19

\bibitem[{{Ilbert} {et~al.}(2010){Ilbert}, {Salvato}, {Le Floc'h}, {Aussel},
  {Capak}, {McCracken}, {Mobasher}, {Kartaltepe}, {Scoville}, {Sanders},
  {Arnouts}, {Bundy}, {Cassata}, {Kneib}, {Koekemoer}, {Le F{\`e}vre}, {Lilly},
  {Surace}, {Taniguchi}, {Tasca}, {Thompson}, {Tresse}, {Zamojski}, {Zamorani},
  \& {Zucca}}]{Ilbert+10}
{Ilbert}, O., {Salvato}, M., {Le Floc'h}, E., {et~al.} 2010, \apj, 709, 644

\bibitem[{{Iovino} {et~al.}(2010){Iovino}, {Cucciati}, {Scodeggio}, {Knobel},
  {Kova{\v c}}, {Lilly}, {Bolzonella}, {Tasca}, {Zamorani}, {Zucca}, {Caputi},
  {Pozzetti}, {Oesch}, {Lamareille}, {Halliday}, {Bardelli}, {Finoguenov},
  {Guzzo}, {Kampczyk}, {Maier}, {Tanaka}, {Vergani}, {Carollo}, {Contini},
  {Kneib}, {Le F{\`e}vre}, {Mainieri}, {Renzini}, {Bongiorno}, {Coppa}, {de la
  Torre}, {de Ravel}, {Franzetti}, {Garilli}, {Le Borgne}, {Le Brun},
  {Mignoli}, {Pell{\`o}}, {Peng}, {Perez-Montero}, {Ricciardelli}, {Silverman},
  {Tresse}, {Abbas}, {Bottini}, {Cappi}, {Cassata}, {Cimatti}, {Koekemoer},
  {Leauthaud}, {Maccagni}, {Marinoni}, {McCracken}, {Memeo}, {Meneux},
  {Porciani}, {Scaramella}, {Schiminovich}, \& {Scoville}}]{iovino+10}
{Iovino}, A., {Cucciati}, O., {Scodeggio}, M., {et~al.} 2010, \aap, 509, A40

\bibitem[{{Kennicutt}(1998)}]{Kennicutt98}
{Kennicutt}, Jr., R.~C. 1998, \araa, 36, 189

\bibitem[{{Kere{\v s}} {et~al.}(2005){Kere{\v s}}, {Katz}, {Weinberg}, \&
  {Dav{\'e}}}]{keres+05}
{Kere{\v s}}, D., {Katz}, N., {Weinberg}, D.~H., \& {Dav{\'e}}, R. 2005,
  \mnras, 363, 2

\bibitem[{{Kim} {et~al.}(1995){Kim}, {Sanders}, {Veilleux}, {Mazzarella}, \&
  {Soifer}}]{kim+95}
{Kim}, D.-C., {Sanders}, D.~B., {Veilleux}, S., {Mazzarella}, J.~M., \&
  {Soifer}, B.~T. 1995, \apjs, 98, 129

\bibitem[{{Kova{\v c}} {et~al.}(2010){Kova{\v c}}, {Lilly}, {Knobel},
  {Bolzonella}, {Iovino}, {Carollo}, {Scarlata}, {Sargent}, {Cucciati},
  {Zamorani}, {Pozzetti}, {Tasca}, {Scodeggio}, {Kampczyk}, {Peng}, {Oesch},
  {Zucca}, {Finoguenov}, {Contini}, {Kneib}, {Le F{\`e}vre}, {Mainieri},
  {Renzini}, {Bardelli}, {Bongiorno}, {Caputi}, {Coppa}, {de la Torre}, {de
  Ravel}, {Franzetti}, {Garilli}, {Lamareille}, {Le Borgne}, {Le Brun},
  {Maier}, {Mignoli}, {Pello}, {Perez Montero}, {Ricciardelli}, {Silverman},
  {Tanaka}, {Tresse}, {Vergani}, {Abbas}, {Bottini}, {Cappi}, {Cassata},
  {Cimatti}, {Fumana}, {Guzzo}, {Koekemoer}, {Leauthaud}, {Maccagni},
  {Marinoni}, {McCracken}, {Memeo}, {Meneux}, {Porciani}, {Scaramella}, \&
  {Scoville}}]{kovac+10}
{Kova{\v c}}, K., {Lilly}, S.~J., {Knobel}, C., {et~al.} 2010, \apj, 718, 86

\bibitem[{{Kurk} {et~al.}(2008){Kurk}, {Cimatti}, {Zamorani}, {Halliday},
  {Mignoli}, {Pozzetti}, {Daddi}, {Rosati}, {Dickinson}, {Bolzonella},
  {Cassata}, {Renzini}, {Franceschini}, {Rodighiero}, \& {Berta}}]{Kurk+08}
{Kurk}, J., {Cimatti}, A., {Zamorani}, G., {et~al.} 2008, in Astronomical
  Society of the Pacific Conference Series, Vol. 399, Panoramic Views of Galaxy
  Formation and Evolution, ed. {T.~Kodama, T.~Yamada, \& K.~Aoki}, 332

\bibitem[{{Le Floc'h} {et~al.}(2009){Le Floc'h}, {Aussel}, {Ilbert},
  {Riguccini}, {Frayer}, {Salvato}, {Arnouts}, {Surace}, {Feruglio},
  {Rodighiero}, {Capak}, {Kartaltepe}, {Heinis}, {Sheth}, {Yan}, {McCracken},
  {Thompson}, {Sanders}, {Scoville}, \& {Koekemoer}}]{LeFloch+09}
{Le Floc'h}, E., {Aussel}, H., {Ilbert}, O., {et~al.} 2009, \apj, 703, 222

\bibitem[{{Le Floc'h} {et~al.}(2005){Le Floc'h}, {Papovich}, {Dole}, {Bell},
  {Lagache}, {Rieke}, {Egami}, {P{\'e}rez-Gonz{\'a}lez}, {Alonso-Herrero},
  {Rieke}, {Blaylock}, {Engelbracht}, {Gordon}, {Hines}, {Misselt}, {Morrison},
  \& {Mould}}]{LeFloch+05}
{Le Floc'h}, E., {Papovich}, C., {Dole}, H., {et~al.} 2005, \apj, 632, 169

\bibitem[{{Leauthaud} {et~al.}(2010){Leauthaud}, {Finoguenov}, {Kneib},
  {Taylor}, {Massey}, {Rhodes}, {Ilbert}, {Bundy}, {Tinker}, {George}, {Capak},
  {Koekemoer}, {Johnston}, {Zhang}, {Cappelluti}, {Ellis}, {Elvis}, {Giodini},
  {Heymans}, {Le F{\`e}vre}, {Lilly}, {McCracken}, {Mellier},
  {R{\'e}fr{\'e}gier}, {Salvato}, {Scoville}, {Smoot}, {Tanaka}, {Van
  Waerbeke}, \& {Wolk}}]{Leauthaud+10}
{Leauthaud}, A., {Finoguenov}, A., {Kneib}, J.-P., {et~al.} 2010, \apj, 709, 97

\bibitem[{{Lilly} {et~al.}(2009){Lilly}, {Le Brun}, {Maier}, {Mainieri},
  {Mignoli}, {Scodeggio}, {Zamorani}, {Carollo}, {Contini}, {Kneib}, {Le
  F{\`e}vre}, {Renzini}, {Bardelli}, {Bolzonella}, {Bongiorno}, {Caputi},
  {Coppa}, {Cucciati}, {de la Torre}, {de Ravel}, {Franzetti}, {Garilli},
  {Iovino}, {Kampczyk}, {Kovac}, {Knobel}, {Lamareille}, {Le Borgne}, {Pello},
  {Peng}, {P{\'e}rez-Montero}, {Ricciardelli}, {Silverman}, {Tanaka}, {Tasca},
  {Tresse}, {Vergani}, {Zucca}, {Ilbert}, {Salvato}, {Oesch}, {Abbas},
  {Bottini}, {Capak}, {Cappi}, {Cassata}, {Cimatti}, {Elvis}, {Fumana},
  {Guzzo}, {Hasinger}, {Koekemoer}, {Leauthaud}, {Maccagni}, {Marinoni},
  {McCracken}, {Memeo}, {Meneux}, {Porciani}, {Pozzetti}, {Sanders},
  {Scaramella}, {Scarlata}, {Scoville}, {Shopbell}, \& {Taniguchi}}]{Lilly+09}
{Lilly}, S.~J., {Le Brun}, V., {Maier}, C., {et~al.} 2009, \apjs, 184, 218

\bibitem[{{Lilly} {et~al.}(2007){Lilly}, {Le F{\`e}vre}, {Renzini}, {Zamorani},
  {Scodeggio}, {Contini}, {Carollo}, {Hasinger}, {Kneib}, {Iovino}, {Le Brun},
  {Maier}, {Mainieri}, {Mignoli}, {Silverman}, {Tasca}, {Bolzonella},
  {Bongiorno}, {Bottini}, {Capak}, {Caputi}, {Cimatti}, {Cucciati}, {Daddi},
  {Feldmann}, {Franzetti}, {Garilli}, {Guzzo}, {Ilbert}, {Kampczyk}, {Kovac},
  {Lamareille}, {Leauthaud}, {Borgne}, {McCracken}, {Marinoni}, {Pello},
  {Ricciardelli}, {Scarlata}, {Vergani}, {Sanders}, {Schinnerer}, {Scoville},
  {Taniguchi}, {Arnouts}, {Aussel}, {Bardelli}, {Brusa}, {Cappi}, {Ciliegi},
  {Finoguenov}, {Foucaud}, {Franceschini}, {Halliday}, {Impey}, {Knobel},
  {Koekemoer}, {Kurk}, {Maccagni}, {Maddox}, {Marano}, {Marconi}, {Meneux},
  {Mobasher}, {Moreau}, {Peacock}, {Porciani}, {Pozzetti}, {Scaramella},
  {Schiminovich}, {Shopbell}, {Smail}, {Thompson}, {Tresse}, {Vettolani},
  {Zanichelli}, \& {Zucca}}]{Lilly+07}
{Lilly}, S.~J., {Le F{\`e}vre}, O., {Renzini}, A., {et~al.} 2007, \apjs, 172,
  70

\bibitem[{{Lutz} {et~al.}(2011){Lutz}, {Poglitsch}, {Altieri}, {Andreani},
  {Aussel}, {Berta}, {Bongiovanni}, {Brisbin}, {Cava}, {Cepa}, {Cimatti},
  {Daddi}, {Dominguez-Sanchez}, {Elbaz}, {F{\"o}rster Schreiber}, {Genzel},
  {Grazian}, {Gruppioni}, {Harwit}, {Le Floc'h}, {Magdis}, {Magnelli},
  {Maiolino}, {Nordon}, {P{\'e}rez Garc{\'{\i}}a}, {Popesso}, {Pozzi},
  {Riguccini}, {Rodighiero}, {Saintonge}, {Sanchez Portal}, {Santini}, {Shao},
  {Sturm}, {Tacconi}, {Valtchanov}, {Wetzstein}, \& {Wieprecht}}]{Lutz+11}
{Lutz}, D., {Poglitsch}, A., {Altieri}, B., {et~al.} 2011, \aap, 532, A90

\bibitem[{{Magliocchetti} {et~al.}(2014){Magliocchetti}, {Lapi}, {Negrello},
  {De Zotti}, \& {Danese}}]{Magliocchetti+14}
{Magliocchetti}, M., {Lapi}, A., {Negrello}, M., {De Zotti}, G., \& {Danese},
  L. 2014, \mnras, 437, 2263

\bibitem[{{Magliocchetti} {et~al.}(2013){Magliocchetti}, {Popesso}, {Rosario},
  {Lutz}, {Aussel}, {Berta}, {Altieri}, {Andreani}, {Cepa}, {Casta{\~n}eda},
  {Cimatti}, {Elbaz}, {Genzel}, {Grazian}, {Gruppioni}, {Ilbert}, {Le Floc'h},
  {Magnelli}, {Maiolino}, {Nordon}, {Poglitsch}, {Pozzi}, {Riguccini},
  {Rodighiero}, {Sanchez-Portal}, {Santini}, {F{\"o}rster Schreiber}, {Sturm},
  {Tacconi}, \& {Valtchanov}}]{Magliocchetti+13}
{Magliocchetti}, M., {Popesso}, P., {Rosario}, D., {et~al.} 2013, \mnras, 433,
  127

\bibitem[{{Magnelli} {et~al.}(2009){Magnelli}, {Elbaz}, {Chary}, {Dickinson},
  {Le Borgne}, {Frayer}, \& {Willmer}}]{magnelli+09}
{Magnelli}, B., {Elbaz}, D., {Chary}, R.~R., {et~al.} 2009, \aap, 496, 57

\bibitem[{{Magnelli} {et~al.}(2011){Magnelli}, {Elbaz}, {Chary}, {Dickinson},
  {Le Borgne}, {Frayer}, \& {Willmer}}]{magnelli+11}
{Magnelli}, B., {Elbaz}, D., {Chary}, R.~R., {et~al.} 2011, \aap, 528, A35

\bibitem[{{Magnelli} {et~al.}(2013){Magnelli}, {Popesso}, {Berta}, {Pozzi},
  {Elbaz}, {Lutz}, {Dickinson}, {Altieri}, {Andreani}, {Aussel},
  {B{\'e}thermin}, {Bongiovanni}, {Cepa}, {Charmandaris}, {Chary}, {Cimatti},
  {Daddi}, {F{\"o}rster Schreiber}, {Genzel}, {Gruppioni}, {Harwit}, {Hwang},
  {Ivison}, {Magdis}, {Maiolino}, {Murphy}, {Nordon}, {Pannella}, {P{\'e}rez
  Garc{\'{\i}}a}, {Poglitsch}, {Rosario}, {Sanchez-Portal}, {Santini}, {Scott},
  {Sturm}, {Tacconi}, \& {Valtchanov}}]{magnelli+13}
{Magnelli}, B., {Popesso}, P., {Berta}, S., {et~al.} 2013, \aap, 553, A132

\bibitem[{{Mamon} {et~al.}(2013){Mamon}, {Biviano}, \& {Bou{\'e}}}]{Mamon+13}
{Mamon}, G.~A., {Biviano}, A., \& {Bou{\'e}}, G. 2013, \mnras, 429, 3079

\bibitem[{{Mei} {et~al.}(2012){Mei}, {Stanford}, {Holden}, {Raichoor},
  {Postman}, {Nakata}, {Finoguenov}, {Ford}, {Illingworth}, {Kodama}, {Rosati},
  {Tanaka}, {Huertas-Company}, {Rettura}, {Shankar}, {Carrasco}, {Demarco},
  {Eisenhardt}, {Jee}, {Koyama}, \& {White}}]{mei+12}
{Mei}, S., {Stanford}, S.~A., {Holden}, B.~P., {et~al.} 2012, \apj, 754, 141

\bibitem[{{Mullaney} {et~al.}(2012){Mullaney}, {Pannella}, {Daddi},
  {Alexander}, {Elbaz}, {Hickox}, {Bournaud}, {Altieri}, {Aussel}, {Coia},
  {Dannerbauer}, {Dasyra}, {Dickinson}, {Hwang}, {Kartaltepe}, {Leiton},
  {Magdis}, {Magnelli}, {Popesso}, {Valtchanov}, {Bauer}, {Brandt}, {Del Moro},
  {Hanish}, {Ivison}, {Juneau}, {Luo}, {Lutz}, {Sargent}, {Scott}, \&
  {Xue}}]{mullaney+12}
{Mullaney}, J.~R., {Pannella}, M., {Daddi}, E., {et~al.} 2012, \mnras, 419, 95

\bibitem[{{Neistein} {et~al.}(2006){Neistein}, {van den Bosch}, \&
  {Dekel}}]{NvdBD06}
{Neistein}, E., {van den Bosch}, F.~C., \& {Dekel}, A. 2006, \mnras, 372, 933

\bibitem[{{Nordon} {et~al.}(2010){Nordon}, {Lutz}, {Shao}, {Magnelli}, {Berta},
  {Altieri}, {Andreani}, {Aussel}, {Bongiovanni}, {Cava}, {Cepa}, {Cimatti},
  {Daddi}, {Dominguez}, {Elbaz}, {F{\"o}rster Schreiber}, {Genzel}, {Grazian},
  {Magdis}, {Maiolino}, {P{\'e}rez Garc{\'{\i}}a}, {Poglitsch}, {Popesso},
  {Pozzi}, {Riguccini}, {Rodighiero}, {Saintonge}, {Sanchez-Portal}, {Santini},
  {Sturm}, {Tacconi}, {Valtchanov}, {Wetzstein}, \& {Wieprecht}}]{nordon+10}
{Nordon}, R., {Lutz}, D., {Shao}, L., {et~al.} 2010, \aap, 518, L24

\bibitem[{{Papovich} {et~al.}(2010){Papovich}, {Momcheva}, {Willmer},
  {Finkelstein}, {Finkelstein}, {Tran}, {Brodwin}, {Dunlop}, {Farrah}, {Khan},
  {Lotz}, {McCarthy}, {McLure}, {Rieke}, {Rudnick}, {Sivanandam}, {Pacaud}, \&
  {Pierre}}]{papovich+10}
{Papovich}, C., {Momcheva}, I., {Willmer}, C.~N.~A., {et~al.} 2010, \apj, 716,
  1503

\bibitem[{{P{\'e}rez-Gonz{\'a}lez} {et~al.}(2005){P{\'e}rez-Gonz{\'a}lez},
  {Rieke}, {Egami}, {Alonso-Herrero}, {Dole}, {Papovich}, {Blaylock}, {Jones},
  {Rieke}, {Rigby}, {Barmby}, {Fazio}, {Huang}, \&
  {Martin}}]{Perez-Gonzalez+2005}
{P{\'e}rez-Gonz{\'a}lez}, P.~G., {Rieke}, G.~H., {Egami}, E., {et~al.} 2005,
  \apj, 630, 82

\bibitem[{{Planck Collaboration} {et~al.}(2013){Planck Collaboration}, {Ade},
  {Aghanim}, {Armitage-Caplan}, {Arnaud}, {Ashdown}, {Atrio-Barandela},
  {Aumont}, {Baccigalupi}, {Banday}, \& et~al.}]{PC+13}
{Planck Collaboration}, {Ade}, P.~A.~R., {Aghanim}, N., {et~al.} 2013, ArXiv
  e-prints

\bibitem[{{Popesso} {et~al.}(2012){Popesso}, {Biviano}, {Rodighiero},
  {Baronchelli}, {Salvato}, {Saintonge}, {Finoguenov}, {Magnelli}, {Gruppioni},
  {Pozzi}, {Lutz}, {Elbaz}, {Altieri}, {Andreani}, {Aussel}, {Berta}, {Capak},
  {Cava}, {Cimatti}, {Coia}, {Daddi}, {Dannerbauer}, {Dickinson}, {Dasyra},
  {Fadda}, {F{\"o}rster Schreiber}, {Genzel}, {Hwang}, {Kartaltepe}, {Ilbert},
  {Le Floch}, {Leiton}, {Magdis}, {Nordon}, {Patel}, {Poglitsch}, {Riguccini},
  {Sanchez Portal}, {Shao}, {Tacconi}, {Tomczak}, {Tran}, \&
  {Valtchanov}}]{popesso+12}
{Popesso}, P., {Biviano}, A., {Rodighiero}, G., {et~al.} 2012, \aap, 537, A58

\bibitem[{{Popesso} {et~al.}(2009){Popesso}, {Dickinson}, {Nonino}, {Vanzella},
  {Daddi}, {Fosbury}, {Kuntschner}, {Mainieri}, {Cristiani}, {Cesarsky},
  {Giavalisco}, {Renzini}, \& {GOODS Team}}]{Popesso+09}
{Popesso}, P., {Dickinson}, M., {Nonino}, M., {et~al.} 2009, \aap, 494, 443

\bibitem[{{Popesso} {et~al.}(2011){Popesso}, {Rodighiero}, {Saintonge},
  {Santini}, {Grazian}, {Lutz}, {Brusa}, {Altieri}, {Andreani}, {Aussel},
  {Berta}, {Bongiovanni}, {Cava}, {Cepa}, {Cimatti}, {Daddi}, {Dominguez},
  {Elbaz}, {F{\"o}rster Schreiber}, {Genzel}, {Gruppioni}, {Magdis},
  {Maiolino}, {Magnelli}, {Nordon}, {P{\'e}rez Garc{\'{\i}}a}, {Poglitsch},
  {Pozzi}, {Riguccini}, {Sanchez-Portal}, {Shao}, {Sturm}, {Tacconi},
  {Valtchanov}, {Wieprecht}, \& {Wetzstein}}]{Popesso+11}
{Popesso}, P., {Rodighiero}, G., {Saintonge}, A., {et~al.} 2011, \aap, 532,
  A145

\bibitem[{{Prescott} {et~al.}(2006){Prescott}, {Impey}, {Cool}, \&
  {Scoville}}]{Prescott+06}
{Prescott}, M.~K.~M., {Impey}, C.~D., {Cool}, R.~J., \& {Scoville}, N.~Z. 2006,
  \apj, 644, 100

\bibitem[{{Reddy} {et~al.}(2008){Reddy}, {Steidel}, {Pettini}, {Adelberger},
  {Shapley}, {Erb}, \& {Dickinson}}]{reddy+08}
{Reddy}, N.~A., {Steidel}, C.~C., {Pettini}, M., {et~al.} 2008, \apjs, 175, 48

\bibitem[{{Rettura} {et~al.}(2010){Rettura}, {Rosati}, {Nonino}, {Fosbury},
  {Gobat}, {Menci}, {Strazzullo}, {Mei}, {Demarco}, \& {Ford}}]{Rettura+10}
{Rettura}, A., {Rosati}, P., {Nonino}, M., {et~al.} 2010, \apj, 709, 512

\bibitem[{{Robotham} {et~al.}(2011){Robotham}, {Norberg}, {Driver}, {Baldry},
  {Bamford}, {Hopkins}, {Liske}, {Loveday}, {Merson}, {Peacock}, {Brough},
  {Cameron}, {Conselice}, {Croom}, {Frenk}, {Gunawardhana}, {Hill}, {Jones},
  {Kelvin}, {Kuijken}, {Nichol}, {Parkinson}, {Pimbblet}, {Phillipps},
  {Popescu}, {Prescott}, {Sharp}, {Sutherland}, {Taylor}, {Thomas}, {Tuffs},
  {van Kampen}, \& {Wijesinghe}}]{Robotham+11}
{Robotham}, A.~S.~G., {Norberg}, P., {Driver}, S.~P., {et~al.} 2011, \mnras,
  416, 2640

\bibitem[{{Rodighiero} {et~al.}(2010){Rodighiero}, {Vaccari}, {Franceschini},
  {Tresse}, {Le Fevre}, {Le Brun}, {Mancini}, {Matute}, {Cimatti}, {Marchetti},
  {Ilbert}, {Arnouts}, {Bolzonella}, {Zucca}, {Bardelli}, {Lonsdale}, {Shupe},
  {Surace}, {Rowan-Robinson}, {Garilli}, {Zamorani}, {Pozzetti}, {Bondi}, {de
  la Torre}, {Vergani}, {Santini}, {Grazian}, \& {Fontana}}]{rodighiero+10a}
{Rodighiero}, G., {Vaccari}, M., {Franceschini}, A., {et~al.} 2010, \aap, 515,
  A8

\bibitem[{{Rosario} {et~al.}(2012){Rosario}, {Santini}, {Lutz}, {Shao},
  {Maiolino}, {Alexander}, {Altieri}, {Andreani}, {Aussel}, {Bauer}, {Berta},
  {Bongiovanni}, {Brandt}, {Brusa}, {Cepa}, {Cimatti}, {Cox}, {Daddi}, {Elbaz},
  {Fontana}, {F{\"o}rster Schreiber}, {Genzel}, {Grazian}, {Le Floch},
  {Magnelli}, {Mainieri}, {Netzer}, {Nordon}, {P{\'e}rez Garcia}, {Poglitsch},
  {Popesso}, {Pozzi}, {Riguccini}, {Rodighiero}, {Salvato}, {Sanchez-Portal},
  {Sturm}, {Tacconi}, {Valtchanov}, \& {Wuyts}}]{rosario+12}
{Rosario}, D.~J., {Santini}, P., {Lutz}, D., {et~al.} 2012, \aap, 545, A45

\bibitem[{{Rykoff} {et~al.}(2012){Rykoff}, {Koester}, {Rozo}, {Annis},
  {Evrard}, {Hansen}, {Hao}, {Johnston}, {McKay}, \& {Wechsler}}]{Rykoff+12}
{Rykoff}, E.~S., {Koester}, B.~P., {Rozo}, E., {et~al.} 2012, \apj, 746, 178

\bibitem[{{Sanders} {et~al.}(2003){Sanders}, {Mazzarella}, {Kim}, {Surace}, \&
  {Soifer}}]{sanders+03}
{Sanders}, D.~B., {Mazzarella}, J.~M., {Kim}, D.-C., {Surace}, J.~A., \&
  {Soifer}, B.~T. 2003, \aj, 126, 1607

\bibitem[{{Sanders} \& {Mirabel}(1996)}]{sandres_mirabel+96}
{Sanders}, D.~B. \& {Mirabel}, I.~F. 1996, \araa, 34, 749

\bibitem[{{Sanders} {et~al.}(2007){Sanders}, {Salvato}, {Aussel}, {Ilbert},
  {Scoville}, {Surace}, {Frayer}, {Sheth}, {Helou}, {Brooke}, {Bhattacharya},
  {Yan}, {Kartaltepe}, {Barnes}, {Blain}, {Calzetti}, {Capak}, {Carilli},
  {Carollo}, {Comastri}, {Daddi}, {Ellis}, {Elvis}, {Fall}, {Franceschini},
  {Giavalisco}, {Hasinger}, {Impey}, {Koekemoer}, {Le F{\`e}vre}, {Lilly},
  {Liu}, {McCracken}, {Mobasher}, {Renzini}, {Rich}, {Schinnerer}, {Shopbell},
  {Taniguchi}, {Thompson}, {Urry}, \& {Williams}}]{Sanders+07}
{Sanders}, D.~B., {Salvato}, M., {Aussel}, H., {et~al.} 2007, \apjs, 172, 86

\bibitem[{{Sanders} {et~al.}(1988){Sanders}, {Soifer}, {Elias}, {Madore},
  {Matthews}, {Neugebauer}, \& {Scoville}}]{sanders+88}
{Sanders}, D.~B., {Soifer}, B.~T., {Elias}, J.~H., {et~al.} 1988, \apj, 325, 74

\bibitem[{{Santini} {et~al.}(2009){Santini}, {Fontana}, {Grazian}, {Salimbeni},
  {Fiore}, {Fontanot}, {Boutsia}, {Castellano}, {Cristiani}, {de Santis},
  {Gallozzi}, {Giallongo}, {Menci}, {Nonino}, {Paris}, {Pentericci}, \&
  {Vanzella}}]{Santini+09}
{Santini}, P., {Fontana}, A., {Grazian}, A., {et~al.} 2009, \aap, 504, 751

\bibitem[{{Santini} {et~al.}(2012){Santini}, {Rosario}, {Shao}, {Lutz},
  {Maiolino}, {Alexander}, {Altieri}, {Andreani}, {Aussel}, {Bauer}, {Berta},
  {Bongiovanni}, {Brandt}, {Brusa}, {Cepa}, {Cimatti}, {Daddi}, {Elbaz},
  {Fontana}, {F{\"o}rster Schreiber}, {Genzel}, {Grazian}, {Le Floc'h},
  {Magnelli}, {Mainieri}, {Nordon}, {P{\'e}rez Garcia}, {Poglitsch}, {Popesso},
  {Pozzi}, {Riguccini}, {Rodighiero}, {Salvato}, {Sanchez-Portal}, {Sturm},
  {Tacconi}, {Valtchanov}, \& {Wuyts}}]{Santini+12}
{Santini}, P., {Rosario}, D.~J., {Shao}, L., {et~al.} 2012, \aap, 540, A109

\bibitem[{{Saunders} {et~al.}(1990){Saunders}, {Rowan-Robinson}, {Lawrence},
  {Efstathiou}, {Kaiser}, {Ellis}, \& {Frenk}}]{saunders+90}
{Saunders}, W., {Rowan-Robinson}, M., {Lawrence}, A., {et~al.} 1990, \mnras,
  242, 318

\bibitem[{{Scoville} {et~al.}(2007){Scoville}, {Capak}, {Giavalisco},
  {Sanders}, {Yan}, {Aussel}, {Ilbert}, {Salvato}, {Mobasher}, \& {Le
  Floc'h}}]{scoville+07}
{Scoville}, N., {Capak}, P., {Giavalisco}, M., {et~al.} 2007, in American
  Institute of Physics Conference Series, Vol. 943, The Science Opportunities
  of the Warm Spitzer Mission Workshop, ed. L.~J. {Storrie-Lombardi} \& N.~A.
  {Silbermann}, 221--228

\bibitem[{{Shao} {et~al.}(2010){Shao}, {Lutz}, {Nordon}, {Maiolino},
  {Alexander}, {Altieri}, {Andreani}, {Aussel}, {Bauer}, {Berta},
  {Bongiovanni}, {Brandt}, {Brusa}, {Cava}, {Cepa}, {Cimatti}, {Daddi},
  {Dominguez-Sanchez}, {Elbaz}, {F{\"o}rster Schreiber}, {Geis}, {Genzel},
  {Grazian}, {Gruppioni}, {Magdis}, {Magnelli}, {Mainieri}, {P{\'e}rez
  Garc{\'{\i}}a}, {Poglitsch}, {Popesso}, {Pozzi}, {Riguccini}, {Rodighiero},
  {Rovilos}, {Saintonge}, {Salvato}, {Sanchez Portal}, {Santini}, {Sturm},
  {Tacconi}, {Valtchanov}, {Wetzstein}, \& {Wieprecht}}]{Shao+10}
{Shao}, L., {Lutz}, D., {Nordon}, R., {et~al.} 2010, \aap, 518, L26

\bibitem[{{Silverman} {et~al.}(2010){Silverman}, {Mainieri}, {Salvato},
  {Hasinger}, {Bergeron}, {Capak}, {Szokoly}, {Finoguenov}, {Gilli}, {Rosati},
  {Tozzi}, {Vignali}, {Alexander}, {Brandt}, {Lehmer}, {Luo}, {Rafferty},
  {Xue}, {Balestra}, {Bauer}, {Brusa}, {Comastri}, {Kartaltepe}, {Koekemoer},
  {Miyaji}, {Schneider}, {Treister}, {Wisotski}, \& {Schramm}}]{silverman+10}
{Silverman}, J.~D., {Mainieri}, V., {Salvato}, M., {et~al.} 2010, \apjs, 191,
  124

\bibitem[{{Simha} {et~al.}(2009){Simha}, {Weinberg}, {Dav{\'e}}, {Gnedin},
  {Katz}, \& {Kere{\v s}}}]{simha+09}
{Simha}, V., {Weinberg}, D.~H., {Dav{\'e}}, R., {et~al.} 2009, \mnras, 399, 650

\bibitem[{{Springel} {et~al.}(2005){Springel}, {White}, {Jenkins}, {Frenk},
  {Yoshida}, {Gao}, {Navarro}, {Thacker}, {Croton}, {Helly}, {Peacock}, {Cole},
  {Thomas}, {Couchman}, {Evrard}, {Colberg}, \& {Pearce}}]{springel+05}
{Springel}, V., {White}, S.~D.~M., {Jenkins}, A., {et~al.} 2005, \nat, 435, 629

\bibitem[{{Tanaka} {et~al.}(2013){Tanaka}, {Finoguenov}, {Mirkazemi}, {Wilman},
  {Mulchaey}, {Ueda}, {Xue}, {Brandt}, \& {Cappelluti}}]{Tanaka+13}
{Tanaka}, M., {Finoguenov}, A., {Mirkazemi}, M., {et~al.} 2013, \pasj, 65, 17

\bibitem[{{Tran} {et~al.}(2009){Tran}, {Saintonge}, {Moustakas}, {Bai},
  {Gonzalez}, {Holden}, {Zaritsky}, \& {Kautsch}}]{Tran+09}
{Tran}, K.-V.~H., {Saintonge}, A., {Moustakas}, J., {et~al.} 2009, \apj, 705,
  809

\bibitem[{{Trump} {et~al.}(2007){Trump}, {Impey}, {McCarthy}, {Elvis},
  {Huchra}, {Brusa}, {Hasinger}, {Schinnerer}, {Capak}, {Lilly}, \&
  {Scoville}}]{Trump+07}
{Trump}, J.~R., {Impey}, C.~D., {McCarthy}, P.~J., {et~al.} 2007, \apjs, 172,
  383

\bibitem[{{van de Voort} {et~al.}(2011){van de Voort}, {Schaye}, {Booth}, \&
  {Dalla Vecchia}}]{voort+11}
{van de Voort}, F., {Schaye}, J., {Booth}, C.~M., \& {Dalla Vecchia}, C. 2011,
  \mnras, 415, 2782

\bibitem[{{Vanzella} {et~al.}(2006){Vanzella}, {Cristiani}, {Dickinson},
  {Kuntschner}, {Nonino}, {Rettura}, {Rosati}, {Vernet}, {Cesarsky},
  {Ferguson}, {Fosbury}, {Giavalisco}, {Grazian}, {Haase}, {Moustakas},
  {Popesso}, {Renzini}, {Stern}, \& {GOODS Team}}]{vanzella+06}
{Vanzella}, E., {Cristiani}, S., {Dickinson}, M., {et~al.} 2006, \aap, 454, 423

\bibitem[{{Veilleux} {et~al.}(2002){Veilleux}, {Kim}, \&
  {Sanders}}]{veilleux+02}
{Veilleux}, S., {Kim}, D.-C., \& {Sanders}, D.~B. 2002, \apjs, 143, 315

\bibitem[{{Wang} {et~al.}(2007){Wang}, {Yang}, {Mo}, \& {van den
  Bosch}}]{wang+07}
{Wang}, Y., {Yang}, X., {Mo}, H.~J., \& {van den Bosch}, F.~C. 2007, \apj, 664,
  608

\bibitem[{{Wetzel} {et~al.}(2013){Wetzel}, {Tinker}, {Conroy}, \& {van den
  Bosch}}]{wetzel+13}
{Wetzel}, A.~R., {Tinker}, J.~L., {Conroy}, C., \& {van den Bosch}, F.~C. 2013,
  \mnras, 432, 336

\bibitem[{{Wolf} {et~al.}(2009){Wolf}, {Arag{\'o}n-Salamanca}, {Balogh},
  {Barden}, {Bell}, {Gray}, {Peng}, {Bacon}, {Barazza}, {B{\"o}hm}, {Caldwell},
  {Gallazzi}, {H{\"a}usler}, {Heymans}, {Jahnke}, {Jogee}, {van Kampen},
  {Lane}, {McIntosh}, {Meisenheimer}, {Papovich}, {S{\'a}nchez}, {Taylor},
  {Wisotzki}, \& {Zheng}}]{wolf+09}
{Wolf}, C., {Arag{\'o}n-Salamanca}, A., {Balogh}, M., {et~al.} 2009, in
  Astronomical Society of the Pacific Conference Series, Vol. 408, The
  Starburst-AGN Connection, ed. W.~{Wang}, Z.~{Yang}, Z.~{Luo}, \& Z.~{Chen},
  248

\bibitem[{{Wuyts} {et~al.}(2011){Wuyts}, {F{\"o}rster Schreiber}, {Lutz},
  {Nordon}, {Berta}, {Altieri}, {Andreani}, {Aussel}, {Bongiovanni}, {Cepa},
  {Cimatti}, {Daddi}, {Elbaz}, {Genzel}, {Koekemoer}, {Magnelli}, {Maiolino},
  {McGrath}, {P{\'e}rez Garc{\'{\i}}a}, {Poglitsch}, {Popesso}, {Pozzi},
  {Sanchez-Portal}, {Sturm}, {Tacconi}, \& {Valtchanov}}]{wuyts+11}
{Wuyts}, S., {F{\"o}rster Schreiber}, N.~M., {Lutz}, D., {et~al.} 2011, \apj,
  738, 106

\bibitem[{{Ziparo} {et~al.}(2013){Ziparo}, {Popesso}, {Biviano}, {Finoguenov},
  {Wuyts}, {Wilman}, {Salvato}, {Tanaka}, {Ilbert}, {Nandra}, {Lutz}, {Elbaz},
  {Dickinson}, {Altieri}, {Aussel}, {Berta}, {Cimatti}, {Fadda}, {Genzel}, {Le
  Flo'ch}, {Magnelli}, {Nordon}, {Poglitsch}, {Pozzi}, {Portal}, {Tacconi},
  {Bauer}, {Brandt}, {Cappelluti}, {Cooper}, \& {Mulchaey}}]{ziparo+13}
{Ziparo}, F., {Popesso}, P., {Biviano}, A., {et~al.} 2013, \mnras, 434, 3089

\bibitem[{{Ziparo} {et~al.}(2014){Ziparo}, {Popesso}, {Finoguenov}, {Biviano},
  {Wuyts}, {Wilman}, {Salvato}, {Tanaka}, {Nandra}, {Lutz}, {Elbaz},
  {Dickinson}, {Altieri}, {Aussel}, {Berta}, {Cimatti}, {Fadda}, {Genzel}, {Le
  Floc'h}, {Magnelli}, {Nordon}, {Poglitsch}, {Pozzi}, {Portal}, {Tacconi},
  {Bauer}, {Brandt}, {Cappelluti}, {Cooper}, \& {Mulchaey}}]{Ziparo+14}
{Ziparo}, F., {Popesso}, P., {Finoguenov}, A., {et~al.} 2014, \mnras, 437, 458

\end{thebibliography}

\end{document}